\documentclass[12pt]{article}
\usepackage{amsmath}
\usepackage{mathrsfs}
\usepackage{enumerate}
\begin{document}
\def\be{\begin{equation}}
\def\bea{\begin{eqnarray}}
\def\ee{\end{equation}}
\def\eea{\end{eqnarray}}
\def\d{\partial}
\def\eps{\varepsilon}
\def\la{\lambda}
\def\b{\bigskip}
\def\nn{\nonumber \\}
\def\p{\partial}
\def\t{\tilde}
\def\h{{1\over 2}}
\def\be{\begin{equation}}
\def\bea{\begin{eqnarray}}
\def\ee{\end{equation}}
\def\eea{\end{eqnarray}}
\def\b{\bigskip}
\def\u{\uparrow}
\def\AA{\mathscr{A}}
\def\DD{\mathscr{D}}
\def\FF{\mathscr{F}}
\def\GG{\mathscr{G}}
\def\LL{\mathscr{L}}
\def\CC{\mathscr{C}}
\def\MM{\mathscr{M}}
\def\PP{\mathscr{P}}
\def\QQ{\mathscr{Q}}
\newcommand{\comment}[2]{#2}

\setcounter{tocdepth}{2}

\makeatletter
\def\blfootnote{\xdef\@thefnmark{}\@footnotetext}  
\makeatother

\begin{center}
{\LARGE Excitations of the Myers--Perry Black Holes}
\\
\vspace{18mm}
{\bf   Oleg Lunin}
\vspace{14mm}

Department of Physics,\\ University at Albany (SUNY),\\ Albany, NY 12222, USA\\ 

\vskip 10 mm

\blfootnote{email: olunin@albany.edu}

\end{center}

\begin{abstract}

We demonstrate separability of the dynamical equations for all $p$--form fluxes in the Myers--Perry--(A)dS geometry, extending the earlier results for electromagnetic field. In the physically important cases of $p=(1,2,3,4)$, we explicitly write the ODEs governing the dynamics of separable solutions.

\b

\end{abstract}

\newpage

\addtocontents{toc}{\vskip-5pt}
\addtocontents{toc}{\protect\enlargethispage{\baselineskip}}
{
\tableofcontents
\addtocontents{toc}{\vskip-5pt}
}

%

\newpage

\section{Introduction}

Black holes are important laboratories for studying classical and quantum gravity. We are entering a new era
in exploration of these objects that was launched by the first direct observations of electromagnetic \cite{BHemObs} and gravitational \cite{BHgravObs} waves produced by them. Reconstruction of images from these measurements required detailed understanding of classical fields in the vicinity of black holes, and the relevant calculations are usually done numerically \cite{NumGR1,NumGR2}. Nevertheless, in the idealized case of an isolated black hole, all excitations can be studied analytically \cite{Teuk}, and these classic results have played an important role in our understanding of classical and quantum properties of the holes. In four dimensions, light excitations of black holes contain four types of fields (scalars, spinors, vectors, and gravitons), and all of them are covered by the Teukolsky's equations \cite{Teuk}. In higher dimensions, one encounters additional excitations, and exploration of their dynamics is the main goal of this article. The motivation for our investigation comes from the desire to understand classical scattering from black holes, as well as quantum effects encoded in the Hawking radiation \cite{Hawking}.

Hawking radiation is one of the most important quantum phenomena in gravitational physics. While the essence of this process is the same for all fields \cite{Hawking}, the detailed wavefunctions of the emitted particles depend on the properties of the underlying dynamical equations. Furthermore, since black holes radiate  all light fields at comparable rates, quantitative understanding of quantum dynamics requires analysis of all available excitations. In four dimensions, wavefunctions for all relevant fields have been computed in \cite{Teuk}.  Although eventually we are interested in explaining four--dimensional phenomena, in the last three decades great insights into physics of black holes have been gained from analyzing these objects in higher dimensions within the framework of string theory \cite{StrVafa,BMPV,5dBH}. In this case light bosonic excitations are not exhausted by scalars, vectors, and symmetric rank-two tensors, but they also contain higher forms predicted by string theory. Detailed study of such fields, which would shed new light on classical and quantum properties of black holes, is the main objective of this article.

\bigskip

Hawking radiation originates from dynamics of quantum fields on nontrivial gravitational backgrounds. For the static black holes, the original calculations for radiation rates and scattering amplitudes have been extended to various fields in higher dimensions \cite{PreAdSCFT}, and important lessons extracted from such investigations inspired formulation of the AdS/CFT correspondence \cite{AdSCFT}. Success of these studies relied on rich 
$SO(d-1)\times U(1)$ isometries of static black holes in $d$ dimensions, which give rise to $(d-1)$ commuting conserved charges and guarantee full separation of variables in equations for all dynamical fields. Unfortunately, rotating black holes, described by the Myers--Perry geometry \cite{MyersPerry}, have only  $[U(1)]^{[(d+1)/2]}$ isometries, which give rise to $[\frac{d+1}{2}]$ conserved quantities, and they are not sufficient for ensuring full integrability of the dynamical equations.
Nevertheless, equations of motion for probe particles and scalar fields in the Myers--Perry geometry turn out to be fully separable due to additional conserved quantities associated with rank-two Killing 
tensors. The hidden symmetries responsible for this separation were first discovered for the Kerr black hole in \cite{Carter1}, and in the last two decades they have been extended to rotating black holes in arbitrary dimensions in \cite{PreKub,Kub1,Kub2,Kub3,KubRev}, including some geometries carrying charges \cite{StrYano,ChLKill}. Remarkably, separation of variables in the wave equation persists for 
the GLPP geometries \cite{GLPP}\footnote{See also \cite{GLPPflwUp} for a general discussion of the GLPP black holes and their properties.}, which generalize the Myers--Perry black holes to solutions with non--zero values of the cosmological constant, so we will briefly discuss excitations of these backgrounds in section \ref{SecAdS}. 

In contrast to the static black holes, where decomposition into spherical harmonics 
persists for fields of arbitrary spin\footnote{While the analysis of spherical harmonics for all $p$--form fields is rather straightforward \cite{StaticPform}, study of gravitational waves is quite challenging even in the backgrounds of static black holes. We refer to \cite{5DGravWave} for the detailed discussion.}, for rotating spacetimes separation of variables for various fields has been worked out  only on a case--by--case basis. In the four--dimensional Kerr geometry, separability of equations for electromagnetic and gravitational waves was proven in the classic work by Teukolsky \cite{Teuk}. Recently separability of Maxwell's equations has been demonstrated for rotating Myers--Perry and GLPP black holes in all dimensions \cite{LMaxw}\footnote{Various extensions of this work, including incorporation of massive vector fields, have been reported in \cite{KubMaxw,KubMaxw1}.}, and the goal of our article is to extend these results to dynamics of higher forms\footnote{To our knowledge, in the past, the dynamics of $p$--forms in black holes geometries has been studied only in static backgrounds \cite{StaticPform}.} which play an important role in string theory. A better understanding of the black hole excitations will give new handles on probing these fascinating objects. 

\bigskip

This paper has the following organization. In section \ref{SecReview} we review the known results pertaining to the Myers--Perry geometry and its excitations. Specifically, we recall the special frames \cite{Kub1,Kub2,Kub3}, which have played crucial role in separating Maxwell's equations in \cite{LMaxw}, and review the procedure for separating variables in the equations for the scalar and vector fields. The notation established in section \ref{SecReview} is used throughout this article. 

In section \ref{Sec2form} we demonstrate separability of the dynamical equation for the two--form, which can be used to describe either NS--NS or Ramond--Ramond fluxes in string theory. We derive the most general separable ansatz in the Myers--Perry geometry and verify that the resulting ODEs contain the expected number of separation constants. In section \ref{Sec3form} these results are extended to the three--form potentials. While equations for the higher forms follow the same pattern, the resulting ODEs become rather complicated, so we just present the general structure in section \ref{SecPform} and write the explicit equation for the four--form in the physically interesting ten--dimensional case in section \ref{Sec4form}. All our results are extended to the GLPP geometry in section \ref{SecAdS}. Some technical details are presented in the appendices.

\section{Summary of the known results}
\label{SecReview}
\renewcommand{\theequation}{2.\arabic{equation}}
\setcounter{equation}{0}

This article is dedicated to analyzing dynamics of various fields in the backgrounds of the Myers--Perry black holes \cite{MyersPerry}, so we begin with reviewing some general properties of these geometries. In particular, section \ref{SecSubMP} establishes the notation used throughout this paper. The main conclusion of this article is that the procedures for solving equations for all $p$--forms follow similar patterns, so in sections \ref{SecMPwave} and \ref{SecMPMaxw} we review the known results for the scalar fields \cite{PreKub,Kub1,Kub2,Kub3,KubRev} and for the one--form potentials \cite{LMaxw}. These constructions will be extended to higher forms in the remaining parts of this article. Readers familiar with the earlier work can go directly to section \ref{Sec2form}.

\subsection{Myers--Perry black hole and its symmetries}
\label{SecSubMP}

Let us review some well-known properties of the $d$--dimensional Myers--Perry geometry \cite{MyersPerry}. The form of metric differs between even and odd values of $d$, so we begin with quoting the solution in $d=2n+2$ dimensions \cite{MyersPerry,Myers}:
\bea\label{MPeven}
ds^2&=&-dt^2+\frac{Mr}{FR}\Big(dt+\sum_{i=1}^n a_i\mu_i^2 d\phi_i\Big)^2+\frac{FR dr^2}{R-Mr}
+\sum_{i=1}^n(r^2+a_i^2)\Big(d\mu_i^2+\mu_i^2 d\phi_i^2\Big)\nn
&&+r^2d\alpha^2.
\eea
Here variables $(\mu_i,\alpha)$ are subject to a constraint
\bea
\alpha^2+\sum_{i=1}^n\mu_i^2=1,
\eea
and functions $F$, $R$ are defined by 
\bea\label{MPdefFR}
F=1-\sum_{k=1}^n\frac{a_k^2\mu_k^2}{r^2+a_k^2},\quad R=\prod_{k=1}^n (r^2+a_k^2).
\eea
To study equations for various fields in the background (\ref{MPeven}), we need to review the symmetries of this geometry, which are encoded the Killing--Yano tensors associated with it  \cite{Kub1,Kub2,Kub3,ChLKill,KubRev}.

\bigskip

Metric (\ref{MPeven}) has $(n+1)$ Killing vectors corresponding to constant shifts of $t$ and $\phi_i$, but these isometries are not sufficient to ensure full separation of the wave equation or dynamical equations for tensor fields. For the wave equation such separation is guaranteed by a family of Killing tensors \cite{Kub1,ChLKill}, while already for the Maxwell field separation requires a more rigid structure associated with Killing--Yano tensors (KYT) \cite{LMaxw}. The general equation for an anti--symmetric KYT \cite{Yano} is 
\bea
\nabla_\mu Y_{\nu_1\dots \nu_p}+\nabla_{\nu_1} Y_{\mu\nu_2\dots \nu_p}=0,
\eea
and for the Myers--Perry black holes (\ref{MPeven}), all Killing--Yano tensors are given by a very elegant formula \cite{Kub1,Kub2,Kub3}
\bea\label{Kub2}
Y^{2(n-k)}=\star\left[\wedge h^k\right].
\eea
Here $h$ is a special two--form that will be written below. As demonstrated in \cite{LMaxw}, the eigenvectors of the antisymmetric tensor $h$ play the central role in separation of Maxwell's equations, and we expect them to be important in the study of higher forms as well. Using the eigenvectors of a symmetric tensor $t_{\mu\nu}=h_{\mu\alpha}{h_\nu}^\alpha$ as frames, one finds very simple expressions for the two--form $h$ and for the metric \cite{Kub1,Kub2,Kub3}:
\bea\label{HInFrame}
h=re^r\wedge e^t+\sum_i x_i e^{x_i}\wedge e^i,\quad 
ds^2=-(e^t)^2+(e^r)^2+\sum_k [(e^{x_k})^2+(e^k)^2].
\eea
The frames were constructed in \cite{Kub1,Kub2,Kub3}, and we will write only the components with upper indices, $e_A^\mu$, using the notation introduced in \cite{ChLKill}:
\bea\label{AllFramesMP}
e_t&=&-\sqrt{\frac{R^2}{FR\Delta}}\left[\d_t-
\sum_k\frac{a_k}{r^2+a_k^2}\d_{\phi_k}\right],\quad 
e_r=\sqrt{\frac{\Delta}{FR}}\d_r,\nn
e_i&=&-\sqrt{\frac{H_i}{d_i}}\left[\d_t-\sum_k\frac{a_k}{a_k^2-x_i^2}\d_{\phi_k}
\right],\quad e_{x_i}=\sqrt{\frac{H_i}{d_i}}\d_{x_i}\,.
\eea
Here we defined convenient expressions
\bea\label{MiscElliptic}
d_i=(r^2+x_i^2)\prod_{k\ne i}(x_k^2-x_i^2),\quad H_i=\prod_k(a_k^2-x_i^2),\quad
\Delta=R-Mr.
\eea
The map between coordinates $\{\mu_i\}$ and $\{x_k\}$ can be found in \cite{ChLKill}, we just recall  that in the variables $(r,x_i)$, functions (\ref{MPdefFR}) become
\bea\label{MisclEllipticEvev}
R=\prod_{k} (r^2+a_k^2),\quad
FR=\prod_k(r^2+x_k^2).
\eea
Note that, apart from the common overall factors, the components $(e_t^\mu,e_r^\mu)$ of the frames depend only on $r$, while the components $(e_i^\mu,e_{x_i}^\mu)$ depend only on $x_i$. This crucial fact is responsible for separation of variables in all equations discussed in this article. 

As found in \cite{LMaxw} and reviewed in section \ref{SecMPMaxw}, separation of the Maxwell's equations occurs not in components $(A_t,A_r,A_{\phi_i},A_{x_i})$, but in projections of the gauge fields to particular combinations of the frames (\ref{AllFramesMP}). Specifically, it is convenient to define $m^{(I)}_\pm$ by
\bea\label{Mframes}
&&m^{(0)}_\pm\equiv \sqrt{FR}(e_r\mp e_t)
=\frac{R}{\sqrt{\Delta}}\left\{\frac{\Delta}{R}\d_r\pm 
\left[\d_t-
\sum_k\frac{a_k}{r^2+a_k^2}\d_{\phi_k}\right]\right\},\nn
&&m_\pm^{(j)}\equiv\sqrt{d_i}(e_{x_i}\mp ie_i)
=\sqrt{{H_j}}\left\{\d_{x_j}\pm i\left[\d_t-\sum_k\frac{a_k}{a_k^2-x_j^2}\d_{\phi_k}
\right]\right\}\,.
\eea
Then, as demonstrated in \cite{LMaxw}, it is the set of projections $[m^{(I)}_\pm]^\mu A_\mu$ that separates. Note that the components $m_\pm^{(j)\mu}$ depend only on $x_j$, and $m_\pm^{(0)\mu}$ depend only on $r$. The metric and the generator $h$ of the Killing--Yano tensors (\ref{Kub2})--(\ref{HInFrame}) can be written as 
\bea\label{MPevenInFrames}
ds^2&=&\frac{1}{FR}{\tilde m}^{(0)}_{+}{\tilde m}^{(0)}_{-}+
\sum_k\frac{1}{d_k}{\tilde m}^{(k)}_{+}{\tilde m}^{(k)}_{-},\quad {\tilde m}^{(I)}_\pm\equiv  [{m}^{(I)}_\pm]_\mu dx^\mu\\
h&=&\frac{r}{2FR}{\tilde m}^{(0)}_+\wedge {\tilde m}^{(0)}_-+\sum_{k=1}^n \frac{x_k}{2id_k}
{\tilde m}^{(k)}_{+}\wedge{\tilde m}^{(k)}_{-}\,.\nonumber
\eea
The frames (\ref{Mframes}) will play an important role throughout this article. 

\bigskip

Let us now discuss the Myers--Perry black holes in odd dimensions ($d=2n+1$). The metric is given by \cite{MyersPerry,Myers}
\bea\label{MPodd}
\hskip -0.3cm
ds^2=-dt^2+\frac{Mr^2}{FR}\Big(dt+\sum_{i=1}^n a_i\mu_i^2 d\phi_i\Big)^2+\frac{FR dr^2}{R-Mr^2}
+\sum_{i=1}^n(r^2+a_i^2)\Big(d\mu_i^2+\mu_i^2 d\phi_i^2\Big),
\eea
and coordinates $\mu_i$ are subject to a constraint
\bea
\sum_{i=1}^n\mu_i^2=1.\nonumber
\eea
The counterparts of the special frames (\ref{HInFrame}) are given by \cite{Kub1,ChLKill}
\bea\label{AllFramesMPOdd}
e_t&=&-\sqrt{\frac{R^2}{FR\Delta}}\left[ \d_t
-\sum_k\frac{a_k}{r^2+a_k^2}\d_{\phi_k}\right],\quad e_r=\sqrt{\frac{\Delta}{FR}}\d_r,\quad
e_{x_i}=\sqrt{\frac{H_i}{x_i^2 d_i}}\d_{x_i}\nn
e_i&=&-\sqrt{\frac{H_i}{x^2_id_i}}\left[\d_t-\sum_k\frac{a_k}{a_k^2-x_i^2}\d_{\phi_k}
\right],\quad 
e_\psi=-\frac{\prod a_i}{r\prod x_k}\left[\d_t-\sum_k\frac{1}{a_k}\d_{\phi_k}
\right].
\eea
Most relations (\ref{MiscElliptic}), (\ref{MisclEllipticEvev}) still hold, with two exceptions:
\bea\label{DelktaOdd}
FR=r^2\prod_k(r^2+x^2_k),\quad \Delta=R-Mr^2.
\eea
The Killing--Yano tensors still have the form (\ref{Kub2})--(\ref{HInFrame}), although the metric acquires an extra term $(e_\psi)^2$, and we refer to \cite{ChLKill} for the detailed discussion. Finally, the light--cone frames $m^{(I)}_\pm$ are given by counterparts of (\ref{Mframes}), and there is an additional frame $n^\mu$:
\bea\label{GenFramesOddD}
&&\hskip -1cm\left[m_\pm^{(0)}\right]^\mu\d_\mu=\frac{R}{\sqrt{\Delta}}\left\{\frac{\Delta}{R}\d_r\pm \left[\d_t-
\sum_k\frac{a_k}{r^2+a_k^2}\d_{\phi_k}\right]\right\},\qquad \Delta=R-Mr^2,\\
&&\hskip -1cm\left[m_\pm^{(j)}\right]^\mu
\d_\mu=\sqrt{{H_j}}\left\{\d_{x_j}\pm i\left[\d_t-\sum_k\frac{a_k}{a_k^2-x_j^2}\d_{\phi_k}
\right]\right\}\,,\quad
n^\mu\d_\mu=\d_t-\sum_k\frac{1}{a_k}\d_{\phi_k}.\nonumber
\eea
To study dynamical equations, we will need the expression for the inverse metric in terms of these frames: 
\bea
g^{\mu\nu}\d_\mu\d_\nu&=&\frac{1}{FR}\left[m_+^{(0)}\right]^\mu 
\left[m_-^{(0)}\right]^\nu\d_\mu\d_\nu+
\sum_{j=1}^n \frac{1}{x_j^2 d_j}{\left[m_+^{(j)}\right]^\mu\left[m_-^{(j)}\right]^\mu\d_\mu\d_\nu}\\
&&+
\left[\frac{\prod a_i}{r\prod x_k}\right]^2n^\mu n^\nu\d_\mu\d_\nu\,.\nonumber
\eea
Note that the components $[m_\pm^{(j)}]^\mu$ depend only on $x_j$, $[m_\pm^{(0)}]^\mu$ depend only on $r$, and $n^\mu$ are constants. This feature is crucial for ensuring separation of variables in the Klein--Gordon equation and in dynamical equations for $p$--form potentials. In this article we are focusing on massless fields, so we begin with reviewing separation of the wave equation.

\subsection{Separation of the wave equation}
\label{SecMPwave}

As we will see in the later sections, dynamics of a $p$--form in the Myers--Perry geometry is governed by a separable scalar function that satisfies a system of ordinary differential equations. To get an inspiration for the resulting ODEs, it is useful to review separation of variables in the wave equation. Our discussion of the scalar field in the Myers--Perry geometry will be very brief, and we refer to \cite{LMaxw} for the details. 

Let us consider the wave equation
\bea\label{WaveEqn}
\frac{1}{\sqrt{-g}}\d_\mu\left[\sqrt{-g}g^{\mu\nu}\d_\nu \Psi\right]=0
\eea
in the background (\ref{MPeven}). Imposing a separable ansatz
\bea\label{WaveEqnTwo}
\Psi=E\Phi(r)\left[\prod X_i(x_i)\right],\qquad E=e^{i\omega t+i\sum n_i\phi_i}\,,
\eea
we conclude that the functions $(\Phi,X_i)$ satisfy a system of differential equations \cite{LMaxw}:
\bea\label{SeparWaveEvenKerr}
&&\frac{d}{dr}\left[\Delta\frac{d\Phi}{dr}\right]+\frac{R^2}{\Delta}
\left[\omega-
\sum_k\frac{a_k n_k}{r^2+a_k^2}\right]^2\Phi-P_{n-1}[-r^2]\Phi=0\,,\\
&&\frac{d}{dx_i}\left[H_i\frac{dX_i}{dx_i}\right]-
{H_i}\left[\omega-
\sum_k\frac{a_k n_k}{a_k^2-x_i^2}\right]^2X_i+P_{n-1}[x_i^2]X_i=0\,.\nonumber
\eea
Here $P_{n-1}[y]$ is an arbitrary polynomial of degree $(n-1)$, and it is crucial that all equations (\ref{SeparWaveEvenKerr}) contain {\it the same} function $P_{n-1}$. 

Note that the separable ansatz (\ref{WaveEqnTwo}) reduces one PDE (\ref{WaveEqn}) to a system of $d=2(n+1)$ ordinary differential equations for function $\Psi$: $(n+1)$ of them come from (\ref{SeparWaveEvenKerr}) upon replacement $(\Phi,X_i)\rightarrow\Psi$, and the remaining $(n+1)$ relations are
\bea
\d_t\Psi=i\omega\Psi,\quad \d_{\phi_k}\Psi=in_k\Psi\,.
\eea
Thus equations (\ref{WaveEqnTwo}), (\ref{SeparWaveEvenKerr}) guarantee full separability of the wave equation, and as expected, the resulting solution depends on $d-1=2n+1$ integration constants:
\label{CountWave}
\begin{enumerate}[(i)]
\item $(n+1)$ parameters $(\omega,n_i)$;
\item $n$ free coefficients of the polynomial $P_{n-1}$.
\end{enumerate}
The separation (\ref{WaveEqnTwo}) also extends to rotating black holes with cosmological constant \cite{Kub1,Kub2,Kub3}, and we refer to \cite{LMaxw} for the counterparts of equations (\ref{SeparWaveEvenKerr}) for that case. Remarkably, at least in some special situations, solutions of the resulting ODEs can be written in terms of Painleve transcendental functions \cite{PnlvOne}. It would be interesting to see whether a similar reduction to Painleve transcendentals occurs for the ODEs encountered in sections \ref{Sec2form}--\ref{SecPform} of our article\footnote{A recent paper \cite{PnlvTwo} found such a reduction for the Teukolsky's equations \cite{Teuk} describing electromagnetic and gravitational waves in the background of the Kerr black hole.}.

\bigskip

We conclude this short subsection by writing the counterpart of equations  (\ref{SeparWaveEvenKerr}) for the geometry (\ref{MPodd}) in $d=2n+1$ dimensions \cite{LMaxw}:
\bea\label{SeparWaveOddKerr}
&&r\frac{d}{dr}\left[\frac{\Delta}{r}\frac{d\Phi}{dr}\right]+\frac{R^2}{\Delta}
\left[\omega-
\sum_k\frac{a_k n_k}{r^2+a_k^2}\right]^2\Phi-P_{n-2}[-r^2]\Phi=0,\nn
\\
&&{x_i}\frac{d}{dx_i}\left[\frac{H_i}{x_i}\frac{dX_i}{dx_i}\right]-
{H_i}\left[\omega-
\sum_k\frac{a_k n_k}{a_k^2-x_i^2}\right]^2X_i+P_{n-2}[-x_i^2]X_i=0\,.\nonumber
\eea
As before, the relations (\ref{WaveEqnTwo}), (\ref{SeparWaveOddKerr}) guarantee full separability of the wave equation, but the counting of $d-1=2n$ free parameters is slightly different:
\begin{enumerate}[(i)]
\item $(n+1)$ parameters $(\omega,n_i)$;
\item $(n-1)$ free coefficients of the polynomial $P_{n-2}$.
\end{enumerate}
In the remainder of this article we will extend the separation (\ref{WaveEqnTwo}), (\ref{SeparWaveEvenKerr}), (\ref{SeparWaveOddKerr}) to the dynamical equations for all $p$--form potentials. We begin with reviewing the $p=1$ case that has been solved in \cite{LMaxw}.

\subsection{Maxwell's equations in the Myers--Perry geometry}
\label{SecMPMaxw}

The first extension of (\ref{WaveEqnTwo}), (\ref{SeparWaveEvenKerr}), (\ref{SeparWaveOddKerr}) beyond the scalar field has been accomplished in \cite{LMaxw}, and the ansatz found in that article was inspired by the classic analysis of the Maxwell field in the four--dimensional Kerr geometry \cite{Teuk}. We refer to \cite{LMaxw} for the detailed discussion of the relation between the higher--dimensional ansatz for the gauge field and the four--dimensional variables introduced by Teukolsky \cite{Teuk}. 
Here we just mention that, in contract to a gauge--invariant formulation used in \cite{Teuk}, separation of variables in higher dimensions occurred in certain projections of the {\it gauge potential}, so it worked only in a specific gauge. As we will see in later sections, similar preferred gauges persist for all $p$--forms, so it is useful to recall the construction of \cite{LMaxw}. It is convenient to separate the discussions of odd and even dimensions. 

\bigskip
\noindent
{\bf Even dimensions}

As demonstrated in \cite{LMaxw}, the most general separable ansatz for the Maxwell field in even dimensions has the form
\bea\label{MaxwAnstz}
[m^{(I)}_\pm]^\nu A_\nu=\mp\frac{i}{x_I\pm\mu}[m^{(I)}_\pm]^\nu\d_\nu \Psi,\quad x_0=-ir,
\eea
where the scalar function $\Psi$ is given by (\ref{WaveEqnTwo}). Relations (\ref{MaxwAnstz}) have a free parameter $\mu$ that will play a role of one of the separation constants. Equations (\ref{MaxwAnstz}) can be easily solved for the gauge potential,
\bea\label{MaxwAnsAlt}
A_\mu={K_\mu}^\nu\d_\nu\Psi,\quad 
{K_\mu}^\nu=-\sum_I\sum_{\alpha=\pm}\frac{i\alpha}{x_I+\alpha\mu}\frac{1}{d_I}
[m^{(I)}_{-\alpha}]_\mu
[m^{(I)}_\alpha]^\nu\,,
\eea
but we find the expression (\ref{MaxwAnstz}) to be more compact and more useful for extending to higher forms. Substitution of the ansatz (\ref{MaxwAnstz}) into Maxwell's equations leads to a system of ODEs \cite{LMaxw}:
\bea\label{MaxwEvenDim}
&&{D_j}\frac{d}{dx}\left[\frac{H_j}{D_j}X_j'\right]+\left\{\frac{2\Lambda}{D_j}-H_j W_j^2-
\Lambda +P_{n-2}[x_j^2] D_j\right\}X_j=0,
\nn
&&{D_r}\frac{d}{dr}\left[\frac{\Delta}{D_r}{\dot\Phi}\right]-\left\{\frac{2\Lambda}{D_r}-
\frac{R^2 W_r^2}{\Delta}-
\Lambda +P_{n-2}[-r^2]  D_r\right\}\Phi=0.
\eea
Various functions appearing in these equations are defined by
\bea\label{WfuncMaxw}
&&\Omega=\omega-\sum\frac{n_ka_k}{\Lambda_k},\quad 
W_j=\omega-\sum\frac{n_k a_k}{a_k^2-x_j^2},\quad W_r=\omega-\sum\frac{n_k a_k}{a_k^2+r^2}\,,\\
&&
D_j=1-\frac{x_j^2}{\mu^2}\,,\quad D_r=1+\frac{r^2}{\mu^2}\,,\quad
\Lambda=\frac{\Omega}{\mu}
\prod \Lambda_k,\quad \Lambda_i=(a_i^2-{\mu^2})\,.
\nonumber
\eea
Note that the wave equation (\ref{SeparWaveEvenKerr}) is also covered by a minor modification of  the system (\ref{MaxwEvenDim}):
\bea\label{EvenDimGenSpinLmb}
\mbox{scalar}:&&D_r=D_j=1,\quad \forall \Lambda,\quad P_{n-2}\rightarrow P_{n-1}.
\eea
This implies that in the process of going from scalar to vector systems (i.e., from (\ref{SeparWaveEvenKerr}) to (\ref{MaxwEvenDim})), one coefficient in the polynomial $P_{n-1}$ is traded for a separation constant 
$\mu$. Thus the vector counterpart of the counting presented on page \pageref{CountWave} is
\begin{enumerate}[(i)]
\item $(n+1)$ parameters $(\omega,n_i)$;
\item $(n-1)$ free coefficients of the polynomial $P_{n-2}$;
\item constant $\mu$.
\end{enumerate}
As expected, the total number of free parameters is still $d-1=2n+1$. Furthermore, as demonstrated in \cite{LMaxw}, the ansatz (\ref{MaxwAnstz}) covers all $d-2$ polarizations of the electromagnetic field: they correspond to different ranges of $\mu$. 

Note that the ansatz (\ref{MaxwAnstz}) fully fixes the gauge since it is not preserved under a gauge transformation. However, after equations (\ref{MaxwAnstz}), (\ref{MaxwEvenDim}) were {\it derived} in 
\cite{LMaxw} as the system describing the most general separable solution of the Maxwell's equations, it was {\it observed} in \cite{KubMaxw} that the resulting potential happened to be in the Lorentz gauge:
\bea\label{LornzGauge}
\nabla_\mu A^\mu=0.
\eea
This observation allowed the authors of \cite{KubMaxw} to develop an elegant shortcut in deriving the system (\ref{MaxwEvenDim}) by imposing the gauge condition (\ref{LornzGauge}) from the beginning. Unfortunately a similar strategy seems to be failing for higher forms\footnote{See Appendix \ref{AppB} for a detailed discussion.}, so the original derivation presented in \cite{LMaxw} appears to be more suitable for the extension to $p$--forms. 

\bigskip
\noindent
{\bf Odd dimensions}

The counterpart of the ansatz (\ref{MaxwAnstz}) for odd dimensions was also derived in \cite{LMaxw}, where it was shown that the most general separable solution of the Maxwell's equations has the form
\bea\label{NewAnstzOdd}
[m^{(I)}_\pm]^\nu A_\nu=\mp\frac{i}{x_I\pm\mu}[m^{(I)}_\pm]^\nu\d_\nu \Psi,\quad
n^\nu A_\nu=-\frac{i}{\mu}n^\nu\d_\nu\Psi\,,
\eea
with $\Psi$ given by (\ref{WaveEqnTwo}). The resulting ordinary differential equations are \cite{LMaxw}
\bea\label{OddMPmaster}
&&\frac{D_j}{x_j}\frac{d}{dx_j}\left[\frac{H_j}{x_j D_j}X_j'\right]+\left\{\frac{2\Lambda}{D_j}-
\frac{H_jW_j^2}{x_j^2}+\frac{\AA D_j}{x_j^2}{\Omega}^2+P_{n-3}[-x_j^2]D_j\right\}=0,\nn
&&\frac{D_r}{r}\frac{d}{dr}\left[\frac{\Delta}{r D_r}{\dot\Phi}\right]+\left\{\frac{2\Lambda}{D_r}+
\frac{R^2W_r^2}{r^2\Delta}-\frac{\AA D_r}{r^2}{\Omega}^2+P_{n-3}[r^2] D_r\right\}\Phi=0.
\eea
Here functions $(W_j,W_r)$ are still given by (\ref{WfuncMaxw}), while the constants 
$(\AA,{\Omega})$ are defined as
\bea\label{AAdef}
\AA=\left[\prod a_k\right]^2,\quad
{\Omega}=\omega-\sum_k \frac{n_k}{a_k}.
\eea
As in even dimensions, one can show that the system (\ref{NewAnstzOdd})--(\ref{OddMPmaster}) describes the most general separable solution of the Maxwell's equations, and it depends on $d-1$ separation parameters and covers $d-2$ independent polarizations \cite{LMaxw}. In the rest of this article we will extend the systems (\ref{MaxwAnstz}), (\ref{MaxwEvenDim}) and (\ref{NewAnstzOdd})--(\ref{OddMPmaster}) to equations governing the dynamics of $p$--form potentials.

\section{Two--form potential in the Myers--Perry geometry}
\label{Sec2form}
\renewcommand{\theequation}{3.\arabic{equation}}
\setcounter{equation}{0}

In the last section we reviewed separation of variables in equations for scalar and vector fields in the Myers--Perry geometry. While in four dimensions such excitations, along with spinors and gravitons, exhaust all interesting modes, for $d>4$ one also encounters higher forms, such as Ramond--Ramond potentials predicted by string theory.  Since the Myers--Perry black hole is a solution of the type II supergravity, it is interesting to study SUGRA excitations of this geometry, and they include  higher forms corresponding to either Ramond--Ramond potentials or the Kalb--Ramond field. In this section we will extend the success in separating equations for the scalar and vector perturbations to equations of motion for a two--form field, and dynamics of higher forms will be discussed in sections \ref{Sec3form}--\ref{SecPform}. To emphasize the analogy with the Maxwell field and with higher forms, we will denote the dynamical variables by $A$, and our analysis is equally applicable to the Kalb--Ramond field $B$ and to the Ramond--Ramond potential $C^{(2)}$. As in the case of electromagnetism, the discussion naturally splits into even-- and odd--dimensional cases, which are covered in subsections \ref{Sec2formEven},  \ref{Sec2formOdd}. 

\subsection{Even dimensions}
\label{Sec2formEven}

Let us consider a two--form potential $A$ that obeys a system of linear differential equations
\bea\label{FieldEqn}
d\star dA=0.
\eea 
Here Hodge dual is taken with respect to the Myers--Perry geometry (\ref{MPeven}). Since the metric has $(n+1)$ isometries corresponding to translations along the $(t,\phi_i)$ coordinates, we can look for solutions in the form
\bea\label{Aform2}
A=\frac{1}{2}e^{i\omega t+\sum n_i\phi_i}{\tilde A}_{\mu\nu}(r,x_i)dx^\mu\wedge dx^\nu\,,
\quad x^\mu=\{r,x_i,t,\phi_i\}.
\eea
As in the case of electromagnetism, we expect that separable solutions are governed by one ``master function'' 
$\Psi$, and now we will determine the relation between the gauge potential and such function. 

We begin with focusing on a special case 
\bea\label{OmZeroMain}
\omega=n_i=0.
\eea
Since the metric has a block form
\bea
g_{\mu\nu}=\left[
\begin{array}{cc}
g_{ab}&0\\
0&g_{ij}
\end{array}
\right],\qquad \{a,b\}=\{t,\phi_i\},\quad \{i,j\}=\{r,x_i\},\quad x_0=-ir,
\eea
and there is no dependence on $\phi^a=\{t,\phi_i\}$, we conclude that the field equations for configurations (\ref{OmZeroMain}) can be divided into three decoupled groups that involve $A^{(1)}$,  $A^{(2)}$, or $A^{(3)}$:
\bea\label{A123}
A^{(1)}=\frac{1}{2}A_{ab}d\phi^a\wedge d\phi^b,\quad 
A^{(2)}=\frac{1}{2}A_{ij}dx^i\wedge dx^j,\quad 
A^{(3)}=A_{ai}d\phi^a\wedge dx^i.
\eea
Thus in the special case (\ref{OmZeroMain}) one would have three independent ``master functions'' 
$(\Psi^{(1)},\Psi^{(2)},\Psi^{(3)})$, but if a configuration (\ref{A123}) is viewed as a limit of a general solution with nontrivial $(\omega,n_i)$, then all $\Psi^{(a)}$ must be the same, so the dynamics is governed by a single scalar $\Psi$:
\bea\label{OnlyOnePsi}
\Psi=\Psi^{(1)}=\Psi^{(2)}=\Psi^{(3)}.
\eea 
It is clear that we are interested only in excitations of this type. For separable solutions of equations 
(\ref{FieldEqn}), function $\Psi$ must have the form
\bea\label{BfieldSpecPsi}
\Psi=\Phi(r)\prod_i^n X_j(x_j)\,.
\eea

All three cases (\ref{A123}) are analyzed in the Appendix \ref{AppA2}, where it is shown that the most general separable solutions in each class are given by\footnote{Labels $(\alpha,\beta)$ take values $\pm$, and the relevant frames are given by (\ref{Mframes}).} 
\bea\label{NonCyclSolnSmry}
&&A^{(1)}_{ij}=(x_i^2-x_j^2)\frac{px_i^2+q}{Q^{(1)}_i}\frac{px_j^2+q}{Q^{(1)}_j}\d_i\d_j\Psi^{(1)},\nn
&&
\sum_{\mu,\nu}[m^{(I)}_\alpha]^\mu [m^{(J)}_\beta]^\nu A^{(2)}_{\mu\nu}=
\alpha\beta\frac{(x^2_I-x^2_J)x_Ix_J}{{Q}^{(2)}_I{Q}^{(2)}_J}
\sum_{i,j}[m^{(I)}_+]^i[m^{(J)}_+]^j \d_i\d_j\Psi^{(2)},\\
&&
\sum_{\mu}[m^{(I)}_\alpha]^\mu A^{(3)}_{\mu j}=
\alpha\frac{(x^2_I-x^2_j)x_I({\tilde p}x_i^2+{\tilde q})}{{Q}^{(3)}_I{Q}^{(3)}_j}
\sum_{i}[m^{(I)}_+]^i\d_i\d_j\Psi^{(3)}\,.\nonumber
\eea
Polynomials $(Q^{(1)}_j,Q^{(2)}_j,Q^{(3)}_j)$ introduced here have the form
\bea
Q^{(a)}_j=a^{(a)} x_j^4+b^{(a)}x_j^2+c^{(a)},\quad x_0=-i r,
\eea
and a priori there are independent sets of coefficients $(a^{(a)},b^{(a)},c^{(a)})$ for three values of the superscript. Configurations (\ref{NonCyclSolnSmry}) solve the field equations (\ref{FieldEqn}) if and only if each of the functions $(\Psi^{(1)},\Psi^{(2)},\Psi^{(3)})$ obeys a system of ODEs
\bea
\d_j\left[\frac{H_j}{Q^{(a)}_j}\d_j\Psi^{(a)}\right]+P^{(a)}_{n-3}[x^2_j]\Psi^{(a)}=0,\quad
\d_r\left[\frac{\Delta}{Q^{(a)}_0}\d_r\Psi^{(a)}\right]-P^{(a)}_{n-3}[-r^2]\Psi^{(a)}=0,
\eea
where $P^{(a)}_{n-3}[x^2_j]$ is a polynomial of degree $(n-3)$ in its argument. Imposing the restriction (\ref{OnlyOnePsi}), we conclude that functions $Q^{(a)}_j$ must be the same for all three values of $a$, and $P^{(a)}_{n-3}$ must have the same property.

Combining relations (\ref{NonCyclSolnSmry}), we find the expression for various projections of the most general gauge potential with a separable function $\Psi$:
\bea\label{BfieldAnswSpecTemp}
\sum_{\mu,\nu}[m^{(I)}_\alpha]^\mu [m^{(J)}_\beta]^\nu A_{\mu\nu}=
\frac{(x^2_I-x^2_J)}{{Q}_I{Q}_J}F(x_I,x_J)
\sum_{i,j}[m^{(I)}_+]^i[m^{(J)}_+]^j \d_i\d_j\Psi\,.
\eea
Here $F$ is a polynomial that is at most quadratic in each of its arguments, and motivated by the form of (\ref{MaxwAnstz}) and (\ref{NonCyclSolnSmry}), we require function $F$ to be separable. This leads to identification of $(p,q)$ with 
$({\tilde p},{\tilde q})$ in (\ref{NonCyclSolnSmry}) and to the final expression for the most general separable ansatz in the special case (\ref{OmZeroMain})\footnote{We used the relation 
$[m^{(I)}_-]^i=[m^{(I)}_+]^i$ to replace 
$[m^{(I)}_+]^i$ by $[m^{(I)}_\alpha]^i$ and $[m^{(J)}_+]^j$ by $[m^{(J)}_\beta]^j$.}:
\bea\label{BfieldAnswSpec}
&&\sum_{\mu,\nu}[m^{(I)}_\alpha]^\mu [m^{(J)}_\beta]^\nu A_{\mu\nu}=
(x^2_I-x^2_J)\frac{V_{I,\alpha}}{Q_I}\frac{V_{J,\beta}}{Q_J}
\sum_{i,j}[m^{(I)}_\alpha]^i[m^{(J)}_\beta]^j \d_i\d_j\Psi\,,\\
&&Q_I=a x_I^4+bx_I^2+c,\quad V_{I,\alpha}=px_I^2+q+\alpha x_I.\nonumber
\eea
Here $\Psi$ is a function with factorized dependence on the coordinates (\ref{BfieldSpecPsi}), that satisfies ordinary differential equations
\bea\label{BfieldODESpec}
\d_j\left[\frac{H_j}{Q_j}\d_j\Psi\right]+P_{n-3}[x^2_j]\Psi=0,\quad
\d_r\left[\frac{\Delta}{Q_0}\d_j\Psi\right]-P_{n-3}[-r^2]\Psi=0\,,
\eea
with {\it the same} polynomial $P_{n-3}$ for all values of $j$. Let us now extend the solution (\ref{BfieldAnswSpec}) --(\ref{BfieldODESpec}) beyond the special case (\ref{OmZeroMain}).

\bigskip

Using the separable ansatz (\ref{MaxwAnstz})  for the Maxwell's equations as a guide, we assume that addition of nontrivial angular dependence does not modify relations (\ref{BfieldAnswSpec}), but results in a new expression for function $\Psi$:
\bea\label{PsiBfield}
\Psi=E\Phi(r)\left[\prod X_i(x_i)\right],\qquad E=e^{i\omega t+i\sum n_i\phi_i}\,.
\eea
In other words, we assume that coefficients $(a,b,c,p,q)$ in (\ref{BfieldAnswSpec}) define a separable ansatz for all values of $(\omega,n_i)$, and the quantum numbers $(\omega,n_i)$ affect only the final ODEs. Specifically, using the scalar and vector cases as guides (see equations (\ref{SeparWaveEvenKerr}), (\ref{MaxwEvenDim})) we impose differential equations
\bea
\d_j\left[\frac{H_j}{Q_j}\d_jX_j\right]+f_j[x^2_j]X_j=0,\quad
\d_r\left[\frac{\Delta}{Q_0}\d_r\Phi\right]-f_r[-r^2]\Phi=0,
\eea
and assume that $(\omega,m_i)$ dependence appears only in functions $f_j[x^2_j]$. For nontrivial angular dependence, such restrictive ansatz leads to inconsistency of the field equations (\ref{FieldEqn}) unless coefficients 
$(a,b,c,p,q)$ are constrained by relations\footnote{The easiest way to see this is to look at $(ij)$ 
components of (\ref{FieldEqn}), but since the relevant expressions are rather unwieldy, we do not quote them here. A detailed derivation of the counterpart of (\ref{abcRatios}) for Maxwell's equations was presented in \cite{LMaxw}, and extension to higher forms follows the same logic.}:
\bea\label{abcRatios}
\frac{a}{c}=\frac{p^2}{q^2},\quad \frac{b}{c}=\frac{2p}{q}-\frac{1}{q^2}\,.
\eea
These restrictions imply factorization of the quartic polynomials $Q_I$:
\bea\label{Qfactor}
Q_I=\frac{a}{p^2}V_{I,+}V_{I,-}\,.
\eea
Once the ansatz (\ref{BfieldAnswSpec}), (\ref{PsiBfield}) with factorization (\ref{Qfactor}) is imposed, all field 
equations (\ref{FieldEqn}) reduce to a system of ODEs, just as in the case of electromagnetism. 

\bigskip

To present the final answer in a compact form, we begin with simplifying the ansatz (\ref{BfieldAnswSpec}) in the presence of the constraint (\ref{Qfactor})\footnote{Recall that $x_0=-ir$.}:
\bea\label{6DanswCrct}
m^{I\mu}_\alpha m^{J\nu}_\beta A_{\mu\nu}=
(x_I^2-x_J^2)\, h^I_\alpha h^J_\beta\, m^{I\mu}_\alpha m^{J\nu}_\beta \d_\mu\d_\nu\Psi,
\quad
h^I_\pm=\frac{1}{e_1+e_2 x_I^2\pm ie_3 x_I}\,.
\eea
Substitution of this configuration with separable function $\Psi$ (\ref{PsiBfield}) into the field 
equations (\ref{FieldEqn}) in $d=2(n+1)$ dimensions leads to a system of ODEs:
\bea\label{EvenBX}
&&D_i\frac{d}{dx_i}\left[\frac{H_i}{D_i}X_i'\right]+\Big[-W_i^2H_i-(e_1-e_2 x_i^2)(G_i-\frac{G_*}{2})+
D_iP_{n-3}[x^2_i]\Big]X_i\nn
&&\qquad\qquad+\frac{ie_3x_i^2}{e_2}X_i\frac{d}{dx_i}\left[\frac{x_i}{x_*^2-x_i^2}
\left\{\frac{H_iW_i}{x_i^2}-
\frac{H_*W_*}{x_*^2}\right\}\right]=0,\\
\label{EvenBR}
&&D_r\frac{d}{dr}\left[\frac{\Delta}{D_r}\Phi'\right]+\Big[\frac{W_r^2 H_r^2}{\Delta}+(e_1+e_2 r^2)(G_r-\frac{G_*}{2})-
D_rP_{n-3}[-r^2]\Big]\Phi\nn
&&\qquad\qquad-\frac{ie_3r^2}{e_2}\Phi\frac{d}{dr}\left[\frac{r}{r^2+x_*^2}
\left\{\frac{H_r W_r}{r^2}+
\frac{H_*W_*}{r_*^2}\right\}\right]=0.
\eea
Here we introduced convenient notation that will be used throughout this article:
\bea\label{GfuncDefine}
&& G_i=\frac{2ie_3H_i}{D_i}W_i,\quad
W_i=\left[\omega-\sum_k\frac{n_k a_k}{a_k^2-x_i^2}\right],\quad 
x_*=i\sqrt{\frac{e_1}{e_2}}\,,\\
&&(G_r,H_r,W_r)=(G_i,H_i,W_i)\Big|_{x_i=ir},\quad (G_\star,H_\star,W_\star)=(G_i,H_i,W_i)\Big|_{x_i=x_*}\,.\nonumber
\eea
We also defined functions $D_j$ and $D_r$ that are specific to the two--form:
\bea\label{Dfunc2form}
D_j=\frac{1}{h^j_+h^j_-},\quad D_r=\frac{1}{h^r_+h^r_-}\,.
\eea
Equations (\ref{6DanswCrct}), (\ref{PsiBfield}), (\ref{EvenBX}), (\ref{EvenBR}) summarize our main result for the two--form gauge potential in the even number of dimensions. As expected, there are $d-1=2n+1$ free separation constants: 
\begin{enumerate}[(i)]
\item
$(n+1)$ parameters $(\omega, n_i)$;
\item 
$(n-2)$ coefficients of the polynomial $P_{n-3}$;
\item two ratios $(e_2/e_1, e_3/e_1)$ appearing in the expressions (\ref{PsiBfield}) for the functions 
$h^I_\pm$. 
\end{enumerate}
Although one can set $e_1=1$ by rescaling the gauge potential, we kept all three parameters 
$(e_1,e_2,e_3)$ to make the expressions more symmetric.  

\bigskip

The counting presented above works for $n\ge 3$, but it clearly fails in four dimensions. 
Formally, for $n=2$ equations (\ref{EvenBX})--(\ref{EvenBR}) contain a polynomial $P_{-2}$ with $(-1)$ degree of freedom\footnote{Note that $P_{-1}$ with zero degrees of freedom should be interpreted as absence of a polynomial. Thus in six dimensions, the separation constants come only from $(\omega, n_i)$ and the ratios $(e_2/e_1, e_3/e_1)$.}, and this clearly indicates the failure of the description (\ref{EvenBX})--(\ref{EvenBR}). It turns out that in the degenerate four--dimensional case, the ratio $e_2/e_1$ is not an independent parameter, but rather it is determined in terms of 
$(\omega, n)$. A direct calculation for the Kerr black holes 
gives
\bea\label{4d2form1}
h^x_\pm=\frac{1}{e_2(x^2-a^2)W_x\pm i\omega e_3 x},\quad 
h^r_\pm=\frac{1}{-e_2(r^2+a^2)W_r\pm \omega e_3 r},\quad {D_x}=\frac{1}{h^x_+ h^x_-}\,.
\eea
The differential equations are
\bea\label{ODE4d2form}
&&\hskip -1cm
D_x\frac{d}{dx}\left[\frac{H_x}{D_x}X'\right]+\left[-W_x^2H_x+\frac{ie_3\omega}{e_2}+
\frac{4ie_2e_3 H_xW_x(\omega x)^2}{D_x}-\frac{2i(e_3\omega)^3 x^2}{e_2D_x}
\right]X=0,\nn
\\
&&\hskip -1cm
D_r\frac{d}{dr}\left[\frac{\Delta}{D_r}\Phi'\right]+\left[\frac{(H_rW_r)^2}{\Delta}
-\frac{ie_3\omega}{e_2}+
\frac{4ie_2e_3 H_rW_r(\omega r)^2}{D_r}-\frac{2i(e_3\omega)^3 r^2}{e_2D_r}
\right]\Phi=0.\nonumber
\eea
Four--dimensional system has another peculiarity: a harmonic two--form $A$ can be dualized into a scalar ${\tilde\Psi}$ using the standard relation
\bea
d{\tilde\Psi}=\star dA.
\eea
A direct calculation shows that a two--form satisfying the ansatz (\ref{6DanswCrct}), (\ref{PsiBfield}), (\ref{4d2form1}) and differential equations (\ref{ODE4d2form}) maps into a separable scalar:
\bea\label{TildePsi4d2form}
{\tilde\Psi}=\frac{4e_3 \omega}{e_2}E{\tilde \Phi}(r){\tilde X}(x),
\eea
where
\bea\label{Dual4dPsi}
{\tilde X}\equiv \frac{1}{D_x}\left[e_2(a^2-x^2)\d_x+ie_3\omega x\right]X,\quad
{\tilde\Phi}\equiv \frac{1}{D_r}\left[e_2(r^2+a^2-Mr)\d_x+e_3\omega r\right]\Phi
\eea
As a consistency check, one can verify that functions $({\tilde X},{\tilde\Phi})$ indeed satisfy the ordinary differential equations (\ref{SeparWaveOddKerr}), which follow from the wave equation for the separable scalar (\ref{TildePsi4d2form}). This concludes our discussion of a two--form field in even dimensions, and in the next subsection our results will be extended to the odd--dimensional case.

\subsection{Odd dimensions}
\label{Sec2formOdd}

The analysis presented in the last subsection can also be carried out for the two--form potential in the odd--dimensional Myers--Perry geometry (\ref{MPodd}). Instead of repeating the algebraic manipulations, we just outline the logical steps:
\begin{enumerate}[(1)]
\label{PageBformLogic}
\item As in section \ref{Sec2formEven} we begin with a special case (\ref{OmZeroMain}) of the general two--form (\ref{Aform2}). Then manipulations presented in the Appendix \ref{AppA2} lead to the most general separable ansatz (\ref{NonCyclSolnSmry}), but now there is a new set of projections: $n^\mu A_{\mu\nu}=0$. Recall that the frames for the metric (\ref{MPodd}) were introduced in (\ref{GenFramesOddD}). 
\item Once the condition (\ref{OmZeroMain}) is relaxed, three types of components (\ref{A123}) become mixed in the equations of motion (\ref{FieldEqn}), so consistency of the separable ansatz (\ref{NonCyclSolnSmry}) requires the identification (\ref{OnlyOnePsi}). Furthermore, by combining various components into counterparts of the relation (\ref{BfieldAnswSpecTemp}), and by requiring separability in the resulting projections, we conclude that the {\it limit} (\ref{OmZeroMain}) of the general separable solution must have the form (\ref{BfieldAnswSpec}): 
\bea\label{BfieldAnswSpecOdd}
&&\sum_{\mu,\nu}[m^{(I)}_\alpha]^\mu [m^{(J)}_\beta]^\nu A_{\mu\nu}=
(x^2_I-x^2_J)\frac{V_{I,\alpha}}{Q_I}\frac{V_{J,\beta}}{Q_J}
\sum_{i,j}[m^{(I)}_\alpha]^i[m^{(J)}_\beta]^j \d_i\d_j\Psi\,.\\
&&n^\mu A_{\mu\nu}=0,\quad Q_I=a x_I^4+bx_I^2+c,\quad V_{I,\alpha}=px_I^2+q+\alpha x_I.\nonumber
\eea
This ansatz leads to an odd--dimensional counterpart of equations (\ref{BfieldODESpec}):
\bea\label{BfieldODESpecOdd}
\frac{1}{x_j}\d_j\left[\frac{H_j}{x_iQ_j}\d_j\Psi\right]+P_{n-4}[x^2_j]\Psi=0,\quad
\frac{1}{r}\d_r\left[\frac{\Delta}{rQ_0}\d_r\Psi\right]+P_{n-4}[-r^2]\Psi=0\,.
\eea
\item Extension of the solution (\ref{BfieldAnswSpecOdd}) beyond the special case (\ref{OmZeroMain}) imposes consistency conditions (\ref{abcRatios}) that lead to  factorization (\ref{Qfactor}). In the odd--dimensional case, introduction of angular dependence also leads to specific nontrivial expressions for 
$n^\mu A_{\mu\nu}$. The final results for the unique separable ansatz and for the ODEs governing the dynamics of the two--form are written below. 
\end{enumerate}
The three steps outlined above lead to the full set of components of the gauge potential
\bea\label{Odd2Anstz}
m^{I\mu}_\alpha m^{J\nu}_\beta A_{\mu\nu}&=&
(x_I^2-x_J^2)\, h^I_\alpha h^J_\beta\, m^{I\mu}_\alpha m^{J\nu}_\beta \d_\mu\d_\nu\Psi,\\
m^{I\mu}_\alpha n^{\nu} A_{\mu\nu}&=&
x_I^2\, h^I_\alpha h^J_\beta\, m^{I\mu}_\alpha n^{\nu}\, \d_\mu\d_\nu\Psi\,,\nonumber
\eea
where
\bea\label{Odd2AnstzP2}
h^I_\pm=\frac{1}{e_1+e_2 x_I^2\pm ie_3 x_I},\quad \Psi=E\Phi(r)\left[\prod X_i(x_i)\right],\qquad E=e^{i\omega t+i\sum n_i\phi_i}\,.
\eea
Various ingredients of function $\Psi$ satisfy the system of ordinary differential equations
\bea\label{OddBX}
&&\hskip -1cm
\frac{D_i}{x_i}\frac{d}{dx_i}\left[\frac{H_i}{x_iD_i}X_i'\right]+\Big[-\frac{W_i^2H_i}{x_i^2}-
2\frac{e_1-e_2 x_i^2}{x_i^2}G_i-
2e_2 G_*+
P_{n-4}[x^2_i]D_i\Big]X_i\nn
&&\qquad+\frac{{\AA}\Omega^2D_i}{e_1^2 x_i^2}X_i-\frac{ie_3x_i}{e_2}X_i\frac{d}{dx_i}\left[\frac{1}{x_i^2-x_*^2}
\left\{\frac{H_iW_i}{x_i^2}-
\frac{H_*W_*}{x_*^2}\right\}\right]=0\,,\nn
\\
&&\hskip -1cm
\frac{D_r}{r}\frac{d}{dr}\left[\frac{\Delta}{rD_r}\Phi'\right]+\Big[\frac{W_r^2H_r^2}{r^2\Delta}+
2\frac{e_1+e_2 r^2}{r^2}G_r-
2e_2 G_*+
P_{n-4}[-r^2]D_r\Big]\Phi\nn
&&\qquad-\frac{{\AA}\Omega^2D_r}{e_1^2 r^2}\Phi+\frac{ie_3r}{e_2}\Phi\frac{d}{dr}\left[\frac{1}{r^2+x_*^2}
\left\{-\frac{H_rW_r}{r^2}-
\frac{H_*W_*}{x_*^2}\right\}\right]=0\,.\nonumber
\eea
Functions appearing in these expressions are given by (\ref{GfuncDefine})--(\ref{Dfunc2form}), and the constants $(\AA,\Omega)$ are defined in (\ref{AAdef}). As expected, solutions of (\ref{OddBX}) contain $d-1=2n$ free separation constants: 
\begin{enumerate}[(i)]
\item
$(n+1)$ parameters $(\omega, n_i)$;
\item 
$(n-3)$ coefficients of the polynomial $P_{n-4}$;
\item two ratios $(e_2/e_1, e_3/e_1)$ appearing in the expressions (\ref{PsiBfield}) for the functions 
$h^I_\pm$. 
\end{enumerate}
Furthermore, as in the case of the Maxwell's equations, different ranges of the ratios $(e_2/e_1, e_3/e_1)$ lead to the correct number of independent polarizations of the two--form. 

\bigskip

The counting presented above fails for $d=5$, which corresponds to $n=2$. In the last subsection we have encountered a similar situation in the four--dimensional case, where the ratios $(e_2/e_1, e_3/e_1)$ were not independent, but rather they were determined by $(\omega,n)$. For the five--dimensional black hole there are two options:
\begin{enumerate}[(a)]
\item The general ansatz (\ref{Odd2Anstz})--(\ref{Odd2AnstzP2}) can be kept, as long as three parameters $(\omega,n_1,n_2)$ are constrained by
\bea\label{OmegaConstr5d}
{\Omega}=\omega-\sum_k \frac{n_k}{a_k}=0.
\eea
Then the differential equations are given by a simpler version of (\ref{OddBX}):
\bea
&&\hskip -2cm
\frac{D_i}{x_i}\frac{d}{dx_i}\left[\frac{H_i}{x_iD_i}X_i'\right]+\Big[-\frac{W_i^2H_i}{x_i^2}-
2\frac{e_1-e_2 x_i^2}{x_i^2}G_i-
2e_2 G_*\Big]X_i=0,\nn
&&\hskip -2cm
\frac{D_r}{r}\frac{d}{dr}\left[\frac{\Delta}{rD_r}\Phi'\right]+\Big[\frac{W_r^2H_r^2}{r^2\Delta}+
2\frac{e_1+e_2 r^2}{r^2}G_r-
2e_2 G_*\Big]\Phi=0.\nonumber
\eea
The second lines in both equations (\ref{OddBX}) disappear due to the constraint (\ref{OmegaConstr5d}) and the explicit form of the product $H_iW_i$ in five dimensions. As expected, there are four separation constants: two ratios $(e_2/e_1, e_3/e_1)$ and two angular momenta $(n_1,n_2)$. The frequency $\omega$ is fixed by the constraint (\ref{OmegaConstr5d}).

\item If the constraint (\ref{OmegaConstr5d}) is not satisfied, then the ansatz (\ref{Odd2AnstzP2}) becomes inconsistent unless $e_2=0$:
\bea\label{5d2form1}
h^I_\pm=\frac{1}{e_1\pm ie_3 x_I}\,.
\eea
The differential equations (\ref{OddBX}) are replaced by
\bea\label{5d2form2}
&&\hskip -2cm
\frac{D_i}{x_i}\frac{d}{dx_i}\left[\frac{H_i}{x_iD_i}X_i'\right]+\Big[-\frac{W_i^2H_i}{x_i^2}+\frac{{\AA}\Omega^2D_i}{e_1^2 x_i^2}+
\frac{2ie_3^3W_\# H_\#}{e_1D_i}\Big]X_i=0,\\
&&\hskip -2cm
\frac{D_r}{r}\frac{d}{dr}\left[\frac{\Delta}{rD_r}\Phi'\right]+\Big[\frac{W_r^2H_r^2}{r^2\Delta}
-\frac{{\AA}\Omega^2D_r}{e_1^2 r^2}+\frac{2ie_3^3W_\# H_\#}{e_1D_r}\Big]\Phi=0.\nonumber
\eea
Here we defined
\bea
(W_\#, H_\#)=(W_x,H_x)|_{x=ie_1/e_3}\,.\nonumber
\eea
Again, there are four separation constants: three unconstrained parameters $(\omega,n_1,n_2)$ and one ratio $e_1/e_3$. 
\end{enumerate}
In five dimensions, a two--form $A^{(2)}$ can be dualized into a vector ${\tilde A}^{(1)}$ using a standard relation
\bea
d{\tilde A}^{(1)}=\star dA^{(2)}.
\eea
The resulting ${\tilde A}^{(1)}$ turns out to have a separable form (\ref{NewAnstzOdd}), and the expression for its ``master function'' ${\tilde\Psi}$ in terms of $\Psi$ is similar to (\ref{TildePsi4d2form})--(\ref{Dual4dPsi}).

\bigskip
\noindent
To summarize, in this section we have demonstrated separability of equations for the two--form potential in all dimensions and found the resulting systems of ODEs that govern the dynamics. In even dimensions, the ansatz and the ODEs are given by (\ref{6DanswCrct}) and (\ref{EvenBX}), while the odd--dimensional results are (\ref{Odd2Anstz}) and (\ref{OddBX}). In the degenerate cases of four and five dimensions, the separable configurations are given by (\ref{4d2form1})--(\ref{ODE4d2form})  and (\ref{5d2form1})--(\ref{5d2form2}). The differential equations contain the correct number of separation constants, and all polarizations are covered by various ranges of the these constants.

\section[Three--form potential in the Myers--Perry geometry]{3--form potential in the Myers--Perry geometry}
\label{Sec3form}
\renewcommand{\theequation}{4.\arabic{equation}}
\setcounter{equation}{0}

In this section we will analyze a three--form potential
\bea
A=\frac{1}{6}A_{\mu\nu\la}dx^\mu\wedge dx^\nu\wedge dx^\la
\eea
and construct the most general separable solution of the equations of motion
\bea\label{3formEqns}
d\star  dA=0. 
\eea
Since the procedure is very similar to the one implemented in the last section, we will only outline the logic and present the results for the most crucial steps. As before, we will have separate discussions for even and odd dimensions.  

\subsection{Even dimensions}

In this subsection we construct the most general separable solution of equations (\ref{3formEqns}) in the Myers--Perry geometry (\ref{MPeven}). Using the results for the Maxwell field and for the two--form as an inspiration, we impose an ansatz
\bea\label{3formAnstz}
m^{I\mu}_\alpha m^{J\nu}_\beta m^{K\la}_\gamma A_{\mu\nu\la}=F[x_I,x_J,x_K]\, h^I_\alpha 
h^J_\beta h^K_\gamma\, m^{I\mu}_\alpha m^{J\nu}_\beta m^{K\la}_\gamma\, \d_\mu\d_\nu\d_\la\Psi,
\eea
with some unknown functions $F$ and $h^I_\pm(x_I)$. Indices $(\alpha,\beta,\gamma)$ in (\ref{3formAnstz}) are not summed over, and they take values $(+,-)$. The frames $m^{I\mu}_\alpha$ are given by (\ref{Mframes}). For separable solutions, function $\Psi$ must have the form
\bea\label{Psi3form}
\Psi=E\Phi(r)\left[\prod X_i(x_i)\right],\qquad E=e^{i\omega t+i\sum n_i\phi_i}\,.
\eea
As in the case of the two--form, we begin with the special case
\bea\label{OmZeroMain3form}
\omega=n_i=0
\eea
to determine the functions $F[x_I,x_J,x_K]$ and $h^I_\alpha$, then we switch on the angular dependence to find the differential equations for $(\Phi,X_j)$. Specifically, we perform the following steps:
\begin{enumerate}[(1)]
\item In the special case (\ref{OmZeroMain3form}), the ansatz (\ref{3formAnstz})--(\ref{Psi3form}) and its counterparts for higher forms are analyzed in Appendix \ref{AppA3}, and the result for the three--form potential reads
\bea\label{FFnc3form}
F[x_I,x_J,x_K]=(x_I^2-x_J^2)(x_I^2-x_K^2)(x_J^2-x_K^2),\quad
h^I_\pm=\frac{p_1x_I^2+p_2\pm p_3 x_I}{Q_I}\,.
\eea
Here polynomials $Q_I$ are defined by
\bea
Q_I=b_3 x_I^6+b_2 x_I^4+b_1x_I^2+b_0,
\eea
and coefficients $(p_k,b_k)$ are the same for all values of $I$. Function $\Psi$ satisfies a system of ordinary differential equations (\ref{SemiFinalODEbP})
\bea\label{3formEqnNcyc}
\d_j\left[\frac{{H_j}}{Q_j}\d_j\right]\Psi+P_{n-4}[x^2_j]\Psi=0,\quad
\d_r\left[\frac{{\Delta}}{Q_0}\d_r\right]\Psi-P_{n-4}[-r^2]\Psi=0,
\eea
where $P_{n-4}[y]$ is an arbitrary polynomial of degree $(n-4)$ in $y$. The system (\ref{FFnc3form})--(\ref{3formEqnNcyc}) gives the {\it most general} separable solution (\ref{3formAnstz}) of equations (\ref{3formEqns}) in the special case (\ref{OmZeroMain3form}).
\item Introduction of angular dependence (i.e., relaxation of the requirements (\ref{OmZeroMain3form})) leads to severe constraints on the coefficients $(p_k,b_k)$. In particular, the roots of two equations 
\bea
p_1x_I^2+p_2+ p_3 x_I=0\quad\mbox{and}\quad p_1x_I^2+p_2- p_3 x_I=0\nonumber
\eea
must also be the roots of $Q_I=0$. A similar situation has already been encountered for the electromagnetic field and for the two--form potential, and in the present case the constraints lead to the final expression for 
$h^I_\pm$:
\bea\label{3formHpm}
h^I_\pm=\frac{1}{[1+e_2 x_I^2\pm ie_3 x_I][1+q x_I^2]}\,.
\eea
This relation is the counterpart of the expression for $h^I_\pm$ from (\ref{6DanswCrct}), although 
now we have rescaled the gauge potential to set $e_1=1$. 
\item In the case of the three--form potential, there are further constraints on parameter $q$, and the separable solution (\ref{3formAnstz})--(\ref{Psi3form}) exists if and only if $q$ satisfies an algebraic equation
\bea\label{qConstraint}
\left[\prod_i(1+q a_i^2)\right]\left[\omega-\sum_i\frac{qa_i n_i}{1+q a_i^2}\right]=0.
\eea
This constraint is automatically satisfied in the special case (\ref{OmZeroMain3form}), but any nontrivial angular dependence leads to $n$ solutions\footnote{Recall that $n$ is related to the dimensions $d$ of the spacetime by $d=2(n+1)$.} for the parameter $q$. Once the constraint (\ref{qConstraint}) is imposed, equations (\ref{3formEqns}) reduce to a system of ODEs (\ref{ODE3form}) for the function $\Psi$. Note that the constraint (\ref{qConstraint}) is a new feature of the three--forms that has not been encountered in $p=1,2$ cases, and in sections \ref{Sec4form}, \ref{SecPform} we will show that the counterparts of the restriction (\ref{qConstraint}) have to be imposed for the $p$--forms with $p>3$ as well.
\end{enumerate}
To summarize, we have found that the most general separable solution of equations (\ref{3formEqns}) has the form
\bea\label{3FrmAnstz}
\hskip -0.4cm
m^{I\mu}_\alpha m^{J\nu}_\beta m^{K\sigma}_\gamma A_{\mu\nu\sigma}=
X_{IJ}X_{IK}X_{JK}\, h^I_\alpha h^J_\beta\, h^K_\gamma
m^{I\mu}_\alpha m^{J\nu}_\beta m^{K\sigma}_\gamma \d_\mu\d_\nu\d_\sigma \Psi,\ \ 
X_{IJ}\equiv x_I^2-x_J^2\,,
\eea
where functions $h^I_\pm$ are given by (\ref{3formHpm}), and parameter $q$ satisfies the constraint (\ref{qConstraint}). The differential equations for functions $(X_j,\Phi)$ are\footnote{To avoid unnecessary clutter, here in similar equations for higher forms we write $X(x)$ instead of $X_j(x_j)$. All functions $X_j$ satisfy the same ODE.} 
\bea\label{ODE3form}
&&\hskip -1cm
D_x\frac{d}{dx}\left[\frac{H_x}{D_x}X'\right]-W_x^2H_xX-
(1+q x^2)(1-e_2 x^2)\left[G_x-\frac{G_\star}{2}\right]X+D_xP_{n-4}[x^2] X\nn
&&\hskip -1cm+\frac{i e_3x^2(1+ qx^2)}{e_2-q}X\frac{d}{dx}\left[
\frac{x(W_x {\hat H}_x-W_\star {\hat H}_\star)}{1+e_2 x^2}-
\frac{x(W_x {\hat H}_x-W_q {\hat H}_q)}{1+q x^2}\right]=0,\\
&&\hskip -1cm
-D_r\frac{d}{dr}\left[\frac{\Delta}{D_r}\Phi'\right]-\frac{W_r^2H_r^2}{\Delta}\Phi-
(1-q r^2)(1+e_2 r^2)\left[G_r-\frac{G_\star}{2}\right]\Phi+D_rP_{n-4}[-r^2] \Phi\nn
&&\hskip -1cm-\frac{i e_3r^2(1-qr^2)}{e_2-q}\Phi\frac{d}{dr}\left[\frac{r(W_r {\hat H_r}-W_\star {\hat H}_\star)}{1-e_2 r^2}-
\frac{r(W_r {\hat H_r}-W_q {\hat H}_q)}{1-q r^2}\right]=0.\nonumber
\eea
Here we used various functions defined in (\ref{GfuncDefine}), as well as some new ingredients:
\bea\label{DxFor3form}
&&{D_x}=\frac{1}{h_+ h_-[1+q x^2]},\quad 
{\hat H}_x=\frac{1}{x^4}\prod(a_i^2-x^2),\quad (D_r,{\hat H}_r)=(D_x,{\hat H}_x)_{x=ir}\nn
&&{\hat H}_\star={e_2^2}\prod(a_i^2+\frac{1}{e_2}),\quad {\hat H}_q={q^2}\prod(a_i^2+\frac{1}{q})
\,.
\eea
This concludes our discussion of three--form potentials in even dimensions.

\subsection{Odd dimensions}
\label{Sec3formOdd}

In odd dimensions, the metric and the frames are given by (\ref{MPodd}), (\ref{GenFramesOddD}). Repeating the analysis presented in the last subsection, we arrive at the most general separable ansatz for the three--form satisfying equations (\ref{3formEqns}):
\bea\label{3FrmAnstzOdd}
m^{I\mu}_\alpha m^{J\nu}_\beta m^{K\sigma}_\gamma A_{\mu\nu\sigma}&=&
X_{IJ}X_{IK}X_{JK}\, h^I_\alpha h^J_\beta\, h^K_\gamma
m^{I\mu}_\alpha m^{J\nu}_\beta m^{K\sigma}_\gamma \d_\mu\d_\nu\d_\sigma \Psi,\quad
X_{IJ}\equiv x_I^2-x_J^2\,,\nn
m^{I\mu}_\alpha m^{J\nu}_\beta n^{\sigma} A_{\mu\nu\sigma}&=&
X_{IJ}x_{I}^2x_{J}^2\, h^I_\alpha h^J_\beta\, 
m^{I\mu}_\alpha m^{J\nu}_\beta n^{\sigma} \d_\mu\d_\nu\d_\sigma \Psi\,.
\eea
Functions $h^I_\alpha$ are still given by (\ref{3formHpm}), where parameter $q$ satisfies an algebraic equation (\ref{qConstraint}). For function $\Psi$ that has the form (\ref{Psi3form}), the dynamical equations
(\ref{3formEqns}) are equivalent to a system of ODEs:
\bea\label{ODE3formOdd}
&&\hskip -1cm
\frac{D_x}{x}\frac{d}{dx}\left[\frac{H_x}{xD_x}X'\right]-\left[\frac{W_x^2H_x}{x^2}+
N_x\left[G_x-\frac{G_\star}{2}\right]-D_xP_{n-5}[x^2]-\frac{{\cal A}\Omega^2D_x}{x^2}\right] X\nn
&&+\left\{
-\frac{ie_2^4H_\star W_\star x^2(1+q x^2)(1+e_2 x^2) }{e_3(e_2-q)}-ie_3e_2(1+q x^2)\FF[W_x H_x]\right\}X=0\,,\nn
&&\phantom{{2}{3}}\\
&&\hskip -1cm
\frac{D_r}{r}\frac{d}{dr}\left[\frac{\Delta}{rD_r}\Phi'\right]-\left[-\frac{W_r^2H_r^2}{r^2\Delta}+
N_r\left[G_r-\frac{G_\star}{2}\right]-D_rP_{n-5}[-r^2]+\frac{{\cal A}\Omega^2D_r}{r^2}\right]\Phi\nn
&&+\left\{
\frac{ie_2^4H_\star W_\star r^2(1-q r^2)(1-e_2 r^2) }{e_3(e_2-q)}-ie_3e_2(1-q r^2)
\Big[\FF[W_x H_x]\Big]_{x=ir}\right\}\Phi=0\,.\nonumber
\eea 
Here we used functions $(H_x,H_r,\Delta,W_x,W_r,G_x,G_r,W_\star,H_\star,G_\star)$ defined in previous sections,
$(D_x,D_r)$ given by (\ref{DxFor3form}), and two new factors $(N_x,N_r)$:
\bea
N_x=(1+q x^2)(1-e_2 x^2)(e^2_2x^2+2e_2+e_3^2),\quad N_r=N_x\Big|_{x=ir}\,.
\eea
Unlike the explicit equations (\ref{EvenBX}), (\ref{OddBX}), (\ref{ODE3form}) encountered earlier, the system (\ref{ODE3formOdd}) contains some unspecified function $\FF$, and unfortunately we were not able to find a closed form expression for $\FF$ that holds in all dimensions. Let us list the properties of $\FF$ which are sufficient for writing equations (\ref{ODE3formOdd}) in all $d\le 11$ that are of interest for supergravity.
\begin{enumerate}[(a)]
\item Function $\FF$ maps polynomials in $x$ into polynomials. 
\item On the space of polynomials, function $\FF$ is linear:
\bea
\FF[\alpha x^m+\beta x^n]=\alpha \FF[x^m]+\beta \FF[x^n].
\eea
\item Property (b) implies that to specify function $\FF$, it is sufficient to evaluate $\FF[x^m]$ for all values of $m$. Furthermore, $H_xW_x$ is a polynomial of degree $\frac{d-1}{2}$ in $x^2$, so for $d\le 11$ we need $\FF[x^m]$ only for the even values of $m\le 10$. The results are\footnote{We also write the result for $m=12$, just to stress that function $\FF$ can be determined even beyond eleven dimensions.}
\bea\label{FFdefine1}
&&\FF[1]=\PP_0-q\PP_2+q^2\PP_4,\quad \FF[x^2]=\PP_2-q\PP_4,\quad 
\FF[x^4]=\PP_4,
\nn
&& \FF[x^6]=0,\quad \FF[x^8]=\frac{1}{q}\PP_6,\quad \FF[x^{10}]=\frac{1}{q}\left[\PP_8-\frac{1}{q}\PP_6\right],\\
&&\FF[x^{12}]=\frac{1}{q}\left[\PP_{10}-\frac{1}{q}\PP_8+\frac{1}{q^2}\PP_6\right]\,.\nonumber
\eea
Here we defined convenient building blocks $\PP_k$:
\bea\label{FFdefine2}
\PP_0&=&e_2 x^{2}+1\,,\quad
\PP_2=x^{2}-\frac{1}{e_2}\,,\quad
\PP_4=2x^{4}-\frac{x^2}{e_2}-\frac{1}{e_2^2}\,,\nn
\PP_6&=&2x^{6}-\frac{2x^4}{e_2}-\frac{3x^2}{e^2_2}+\frac{1}{e_2^3}\,,\quad
\PP_8=2y^{8}-\frac{2x^6}{e_2}-\frac{4x^4}{e^2_2}+\frac{3x^2}{e_2^3}-\frac{1}{e_2^4}\,,\\
\PP_{10}&=&2x^{10}-\frac{2x^8}{e_2}-\frac{6x^6}{e^2_2}+\frac{4x^4}{e_2^3}-\frac{3x^2}{e_2^4}+\frac{1}{e_2^5}\,.\nonumber
\eea
\end{enumerate}
Note that analogues of function $\FF$ have appeared in equations (\ref{EvenBX}), (\ref{OddBX}), (\ref{ODE3form}), but we found closed--form expressions for them. For example, the first equation in (\ref{ODE3form}) can be rewritten as
\bea
&&\hskip -1cm
D_x\frac{d}{dx}\left[\frac{H_x}{D_x}X'\right]-W_x^2H_xX-
(1+q x^2)(1-e_2 x^2)\left[G_x-\frac{G_\star}{2}\right]X+D_xP_{n-3}[x^2] X\nn
&&+\frac{i e_3x^2(1+ qx^2)}{e_2-q}{\tilde\FF}[x]X=0,\nonumber
\eea
where function ${\tilde\FF}$ is defined by 
\bea\label{FFtilde}
{\tilde\FF}[x]=
\frac{d}{dx}\left[
\frac{x(W_x {\hat H}_x-W_\star {\hat H}_\star)}{1+e_2 x^2}-
\frac{x(W_x {\hat H}_x-W_q {\hat H}_q)}{1+q x^2}\right]\,.
\eea
This ${\tilde\FF}$ satisfies the properties (a)--(b) listed above. Unfortunately we were not able to find a simple counterpart of (\ref{FFtilde}) for the function $\FF$ appearing in (\ref{ODE3formOdd}), so we have to rely on the case--by--case construction listed in the item (c). Relations (\ref{FFdefine1})--(\ref{FFdefine2}) are sufficient for writing the ODEs (\ref{ODE3formOdd}) in $d\le 13$ dimensions.

Although equations (\ref{3FrmAnstzOdd}) give the most compact form of the ODEs corresponding to  a separable three--form in odd dimensions, it is convenient to rearrange some terms in (\ref{3FrmAnstzOdd}) to inspire the extension to higher forms discussed in section \ref{SecPform}. It is sufficient to look only at the alternative form of the equation for $X$:
\bea\label{AptP3odd}
&&\hskip -1cm
\frac{D_x}{x}\frac{d}{dx}\left[\frac{H_x}{xD_x}X'\right]+\left[-\frac{W_x^2H_x}{x^2}+
(1+q x^2)\left[e_2 G_\star+\frac{1-e_2 x^2}{x^2}G_x\right]+D_xP_{n-3}[x^2]\right] X\nn
&&+\left[\frac{{\AA}\Omega^2D_x}{x^2}
-\frac{2ie_3H_xW_x(1-e_2x^2)}{x^2}\right]X-ie_3(1+q x^2)\GG[W_x H_x]X=0
\eea
Here $\GG$ is defined a linear map between polynomials of $x$:
\bea
&&\GG[W_x H_x]\equiv\left\{
\frac{e_2^2H_\star W_\star x^2(1-e_2 x^2) }{(e_2-q)}+e_2\FF[W_x H_x]\right\}\,,\nonumber
\eea 
and properties of $\GG$ are similar to those of $\FF$. In section \ref{SecPform} we will propose a conjecture for an extension of equation (\ref{AptP3odd}) to all $p$--forms in odd dimensions. 

\bigskip
\noindent
To summarize, in this section we have demonstrated separability of equations for the three--form potential. In even dimensions, the most general separable ansatz is given by (\ref{3FrmAnstz}), and it leads to the system of ODEs (\ref{ODE3form}). In odd dimensions the results are (\ref{3FrmAnstzOdd}) and (\ref{ODE3formOdd}).

\section{Four--form potential in ten dimensions}
\label{Sec4form}
\renewcommand{\theequation}{5.\arabic{equation}}
\setcounter{equation}{0}

The main motivation for studying $p$--form potentials comes from string theory, which contains dynamical fields with $p\le 4$. In previous sections we have analyzed equations of motion for $p=(1,2,3)$, and we will now study the last important case: the four--form potential. Unfortunately, differential equations describing separable configurations of this field in arbitrary dimensions are rather complicated, so in this section we will focus only on the physically relevant situation of ten dimensions and find the full answer for that case. The structure of equations for the four--potential in arbitrary dimensions will be discussed in the next section, which will also cover all $p$--forms with $p>4$. 

\bigskip

A four--form potential appears in the ten--dimensional type IIB supergravity \cite{IIBSUGRA}, so our analysis focuses on space--times of the form
\bea
\mbox{MP}_d\times T^{10-d}\,,
\eea  
where $\mbox{MP}_d$ is a $d$--dimensional Myers--Perry geometry, and $T^{10-d}$ is a torus. The Ramond--Ramond potential $C^{(4)}$ has four types of components,
\bea\label{Jun20}
C^{(4)}_{\mu\nu\la\sigma},\quad C^{(4)}_{\mu\nu\la a},\quad C^{(4)}_{\mu\nu ab},\quad 
C^{(4)}_{\mu abc},\quad 
C^{(4)}_{abcd},
\eea
where Greek indices correspond to the directions on $\mbox{MP}_d$, and Latin indices cover the torus. Only the first ingredient in (\ref{Jun20}) describes a genuine 4--form on $\mbox{MP}_d$, while other sectors reduce to lower forms. Furthermore, unless $d=10$,  the field
$C^{(4)}_{\mu\nu\la\sigma}$ can be dualized to a four--potential that has at least one index on the torus, i.e., to one of the last three sectors in (\ref{Jun20}). Therefore, equations for the four--potential become interesting only for $d=10$, and here we will focus only on that case. 

In the ten--dimensional type IIB supergravity \cite{IIBSUGRA}, the five--form field strength $F_5$ satisfies the self--duality relation and the Bianchi identity\footnote{We assume that all three--form fluxes are turned off.}
\bea
F_5=\star F_5,\quad dF_5=0.
\eea
To construct separable solutions of these equations using the techniques developed in the previous sections, we begin with the Maxwell--type equation for $C^{(4)}\equiv A^{(4)}$ and build the five--form by adding an exact form $d A^{(4)}$ and its dual\footnote{Here and below, the Ramond--Ramond four--form will be denoted by $A^{(4)}$ rather than $C^{(4)}$ to agree with notation adopted throughout this article.}:
\bea\label{MaxwC4}
&&d\star d A^{(4)}=0,\\ 
&&F_5=d A^{(4)}+\star d A^{(4)}\,.
\eea
In this section we will focus on solving the first equation. 

Using the constructions (\ref{MaxwAnstz}), (\ref{6DanswCrct}), (\ref{3FrmAnstz}) as inspirations, we impose a separable ansatz
\bea\label{4FrmAnstz}
m^{I\mu}_\alpha m^{J\nu}_\beta m^{K\la}_\gamma m^{L\sigma}_\delta A_{\mu\nu\la\sigma}=
F_{IJKL}\, (h^I_\alpha h^J_\beta\, h^K_\gamma h^L_\delta)\,
m^{I\mu}_\alpha m^{J\nu}_\beta m^{K\la}_\gamma m^{L\sigma}_\delta \d_\mu\d_\nu\d_\la\d_\sigma \Psi\,,
\eea
where function $\Psi$ has the form
\bea\label{Psi4form}
\Psi=E\Phi(r)\left[\prod X_i(x_i)\right],\qquad E=e^{i\omega t+i\sum n_i\phi_i}\,.
\eea
In the special case (\ref{OmZeroMain3form}), the resulting equations (\ref{MaxwC4}) are analyzed in the Appendix \ref{AppA3}, where it is shown that, for consistency, function $F_{IJKL}$ must have the form
\bea\label{FFnc4form}
F_{IJKL}=X_{IJ}X_{IK}X_{IL}X_{JK}X_{JL}X_{KL},\quad
X_{IJ}\equiv x_I^2-x_J^2,
\eea
and that factors $h^I_\pm$ must be given by
\bea\label{4formHpnSpec}
h^I_\pm=\frac{p_1x_I^2+p_2\pm p_3 x_I}{Q_I},\quad Q_I=b_4 x_I^8+b_3 x_I^6+b_2 x_I^4+b_1x_I^2+b_0\,.
\eea
Then the Maxwell--type equations (\ref{MaxwC4}) reduce to a system of ODEs:
\bea\label{ODE4formSpec}
\frac{d}{dx_j}\left[\frac{H_j}{Q_j}X_j'\right]=0,\qquad
\frac{d}{dr}\left[\frac{\Delta}{Q_r}\Phi'\right]=0\,.
\eea
As in the case of the lower forms, extension of the ansatz (\ref{4FrmAnstz})--(\ref{4formHpnSpec}) beyond the special configurations (\ref{OmZeroMain3form}) leads to certain restrictions on coefficients $b_k$ in (\ref{4formHpnSpec}). Specifically, we find that factors $h^I_\pm$ must have the form\footnote{To simplify the formulas below, we used the freedom in rescaling $A^{(4)}$ to require $h^I_\pm[0]=1$. An analogous choice in (\ref{6DanswCrct}) would have set $e_1=1$.}
\bea\label{4formHpm}
h^I_\pm=\frac{1}{[1+e_2 x_I^2\pm ie_3 x_I][1+q_1 x_I^2][1+q_2 x_I^2]}\,,
\eea
where $(q_1,q_2)$ are two different roots of the algebraic equation
\bea
\left[\prod_i(1+q a_i^2)\right]\left[\omega-\sum_i\frac{qa_i n_i}{1+q a_i^2}\right]=0\,.
\eea
Note that the same equation (\ref{qConstraint}) has been encountered in our discussion of three--forms. Once the ansatz (\ref{4FrmAnstz}) with functions (\ref{FFnc4form}) and (\ref{4formHpm}) is imposed, equation (\ref{MaxwC4}) reduces to a system of ODEs\footnote{As in previous sections, we write $X(x)$ instead of $X_j(x_j)$ to avoid unnecessary clutter.}
\bea\label{ODE4form}
&&\hskip -1cm
D_x\frac{d}{dx}\left[\frac{H_x}{D_x}X'\right]-W_x^2H_xX-
F_x(1-e_2 x^2)\left[G_x-\frac{G_\star}{2}\right]X
+\frac{ie_3\omega}{e_2q_1q_2}x^2 F_xX=0,\nn
\ \\
&&\hskip -1cm
-D_r\frac{d}{dr}\left[\frac{\Delta}{D_r}\Phi'\right]-\frac{W_r^2H_r^2}{\Delta}\Phi-
F_r(1+e_2 r^2)\left[G_r-\frac{G_\star}{2}\right]\Phi-\frac{ie_3\omega}{e_2q_1q_2}r^2F_r\Phi=0.\nonumber
\eea
Here functions $(G_x,G_r,G_\star)$ are given by (\ref{GfuncDefine}), and functions $(F_x,D_x,F_r,D_r)$ are defined by
\bea
F_x=(1+q_1 x^2)(1+q_2 x^2),\quad 
{D_x}=\frac{1}{h_+ h_- F_x},\quad 
(F_r,D_r)=(F_x,D_x)|_{x=ir}\,.
\eea
Equations (\ref{4FrmAnstz})--(\ref{FFnc4form}), (\ref{4formHpm}), (\ref{ODE4form}) give the full separable solution of the equation (\ref{MaxwC4}) for the four--form in ten dimensions. 

\bigskip

Although solving equation $d\star d A^{(4)}=0$ in arbitrary dimensions is beyond the scope of this paper, the experience with lower forms gained in previous sections suggests that the ansatz  (\ref{4FrmAnstz}), (\ref{FFnc4form}), (\ref{4formHpm}) should hold for all $d\ge 10$. Furthermore, by comparing equations  (\ref{ODE4form}) with the structure of the systems (\ref{MaxwEvenDim}), (\ref{EvenBX}), and (\ref{ODE3form}), we arrive at a reasonable guess for the extension of (\ref{ODE4form}) to all even dimensions $d=2(n+1)$:
\bea\label{ODE4formConj}
&&\hskip -1cm
D_x\frac{d}{dx}\left[\frac{H_x}{D_x}X'\right]-W_x^2H_xX-
F_x\left\{(1-e_2 x^2)\left[G_x-\frac{G_\star}{2}\right]+{ie_3 x^2}{\GG}[W_xH_x]\right\}X\nn
&&+D_xP_{n-5}[x^2] X=0,\nn
&&\phantom{42}\\
&&\hskip -1cm
-D_r\frac{d}{dr}\left[\frac{\Delta}{D_r}\Phi'\right]-\frac{W_r^2H_r^2}{\Delta}\Phi-
F_r\left\{(1+e_2 r^2)\left[G_r-\frac{G_\star}{2}\right]-{ie_3 r^2}{\GG}[W_xH_x]_{x=ir}\right\}\Phi\nn
&&+D_xP_{n-5}[-r^2] \Phi=0.\nonumber
\eea
Function ${\GG}$ is a linear operator on the space of polynomials\footnote{In other words, ${\GG}$ maps polynomials into polynomials and $\GG[\alpha x^m+\beta x^n]=\alpha \GG[x^m]+\beta \GG[x^n]$.}  of $x$, and it will be discussed in more detail in the next section. Equations (\ref{ODE4form}) clearly fit the pattern (\ref{ODE4formConj}) with a specific function $\GG$, and finding the expression for $\GG$ in arbitrary dimensions is an interesting open problem. 

\bigskip
\noindent
To summarize, in this section we have demonstrated separability of the equation for the four--form potential and derived the most general separable ansatz  (\ref{4FrmAnstz}), (\ref{FFnc4form}), (\ref{4formHpm}). In the ten--dimensional case, we also found the system of ODEs  (\ref{ODE4form}) that governs the dynamics, and we conjectured an extension of these equations to all dimensions (\ref{ODE4formConj}). In the next section we will discuss additional conjectures for the ODEs describing $p$--forms with $p>4$.

\section{Separability of equations for higher forms}
\label{SecPform}
\renewcommand{\theequation}{6.\arabic{equation}}
\setcounter{equation}{0}

Although supergravity contains only $p$--forms with $p\le 4$, the construction presented in this article can be extended to cover dynamics of all $p$--forms in all dimensions. In the Appendix \ref{AppA3} we demonstrate that equations for such fields in the Myers--Perry geometry are separable, and in this section we summarize the resulting ansatz. We also propose a conjecture for the system of ordinary differential equations governing the dynamics of $p$--form in all dimensions. 

\bigskip 

Let us consider a dynamical equation for a $p$--form field $A^{(p)}$,
\bea\label{EqnCp}
d\star d A^{(p)}=0,
\eea
in the even--dimensional Myers--Perry geometry (\ref{MPeven}). Inspired by the results of the previous sections, we impose an ansatz
\bea\label{pformAnstz}
m^{I_1\mu_1}_{\alpha_1}\dots m^{I_p\mu_p}_{\alpha_p} A_{\mu_1\dots \mu_p}=F_{I_1\dots I_p} 
\left[\prod h^{I_k}_{\alpha_k}\right]m^{I_1\mu_1}_{\alpha_1}\dots m^{I_p\mu_p}_{\alpha_p}
\d_{\mu_1}\dots \d_{\mu_p}\Psi,
\eea
where
\bea\label{PsiPform}
\Psi=E\Phi(r)\left[\prod X_i(x_i)\right],\qquad E=e^{i\omega t+i\sum n_i\phi_i}\,.
\eea
In the Appendix \ref{AppA3} we analyze the resulting equations (\ref{EqnCp}) in the special case 
\bea\label{OmZeroPform}
\omega=n_i=0.
\eea
and demonstrate that the ansatz (\ref{pformAnstz})--(\ref{PsiPform}) is consistent if and only if\footnote{We focus on $d\ge 2(p+1)$, but lower dimensions admit additional degenerate solutions. Equations 
(\ref{TotalPformSpec}) summarize the final result (\ref{TotalPolBp})--(\ref{TotalPolHHF}) of the Appendix \ref{AppA3}.}
\bea\label{TotalPformSpec}  
F_{I_1\dots I_p}=\Big[\prod^p_{k<l}[x_{I_k}^2-x_{I_l}^2]\Big],\quad 
h_\pm^J=\frac{g_0+g_1 x^2_J\pm g_2x_J}{Q_J}\,,\quad 
Q_J=\sum_{k=0}^p b_k x_J^{2k}\,.
\eea
With this choice, the dynamical equations (\ref{EqnCp}) reduce to a system of ODEs
\bea\label{ODEallFormSpec}
\frac{d}{dx_j}\left[\frac{H_j}{Q_j}X_j'\right]+P_{n-p-1}[x_j^2] X_j=0,\quad
\frac{d}{dr}\left[\frac{\Delta}{Q_r}\Phi'\right]-P_{n-p-1}[-r^2] \Phi=0.
\eea
Here $P_{n-p-1}[y]$ is an arbitrary polynomial of degree $(n-p-1)$ in its argument, and it must be 
the same in all equations (\ref{ODEallFormSpec}). 

Relaxation of the condition (\ref{OmZeroPform}) leads to a complicated set of equations, and a full analysis of this system is beyond the scope of this article. However, by comparing the answers (\ref{MaxwAnstz}), (\ref{6DanswCrct}), (\ref{3formHpm}), (\ref{4formHpm}) for $p=1,2,3,4$, we conjecture that the ansatz (\ref{pformAnstz})--(\ref{PsiPform}) still works, but relations (\ref{TotalPformSpec}) must be replaced by more restrictive requirements\footnote{As in (\ref{3formHpm}) and (\ref{4formHpm}), we normalized the 
$p$--form potential to ensure that 
$h^I_\pm[0]=1$. This avoids unnecessary complications in differential equations (\ref{ODEallFormConj}).}
\bea\label{TotalPformGen}
F_{I_1\dots I_p}=\Big[\prod^p_{k<l}[x_{I_k}^2-x_{I_l}^2]\Big],\quad 
h^I_\pm=\frac{1}{[1+e_2 x_I^2\pm ie_3 x_I]F_I^{(p)}}\,,\quad
F_I^{(p)}=\prod_{k=1}^{p-2}[1+q_k x_I^2]\,.
\eea
Here $\{q_k\}$ are $(p-2)$ different roots of the algebraic equation
\bea\label{EqnForQgen}
\left[\prod_i(1+q a_i^2)\right]\left[\omega-\sum_i\frac{qa_i n_i}{1+q a_i^2}\right]=0\,.
\eea
Comparison of the systems  (\ref{MaxwEvenDim}), (\ref{EvenBX}), (\ref{ODE3form}) leads to the conjecture for the ODEs governing the dynamics  (\ref{EqnCp}) of the $p$--form:
\bea\label{ODEallFormConj}
&&\hskip -1cm
D_x\frac{d}{dx}\left[\frac{H_x}{D_x}X'\right]-W_x^2H_xX-
F^{(p)}_x\left\{(1-e_2 x^2)\left[G_x-\frac{G_\star}{2}\right]+{ie_3 x^2}{\GG}_{p}[W_xH_x]\right\}X\nn
&&+D_xP_{n-p-1}[x^2] X=0,\nn
&&\phantom{42}\\
&&\hskip -1cm
-D_r\frac{d}{dr}\left[\frac{\Delta}{D_r}\Phi'\right]-\frac{W_r^2H_r^2}{\Delta}\Phi-
F^{(p)}_r\left\{(1+e_2 r^2)\left[G_r-\frac{G_\star}{2}\right]-{ie_3 r^2}{\GG}_{p}[W_xH_x]_{x=ir}\right\}\Phi\nn
&&+D_xP_{n-p-1}[-r^2] \Phi=0.\nonumber
\eea
Here we used the expressions (\ref{GfuncDefine}) and introduced factors $(D_x,D_r)$:
\bea
{D_x}=\frac{1}{h^x_+ h^x_- F^{(p)}_x},\quad 
{D_r}=\frac{1}{h^r_+ h^r_- F^{(p)}_r}=D_x|_{x=ir}\,.
\eea
Equations (\ref{ODEallFormConj}) contain a function ${\GG}_{p}$, which depends on $p$, but {\it not} on $n$, and has the following properties:
\begin{enumerate}[(a)]
\item Function $\GG_p$ maps polynomials in $x$ into polynomials. 
\item On the space of polynomials, function $\GG$ is linear:
\bea
\GG_p[\alpha x^m+\beta x^n]=\alpha\, \GG_p[x^m]+\beta\, \GG_p[x^n].\nonumber
\eea
\item Coefficients of the polynomial $\GG_p[x^m]$ depend only on parameters $(e_2,q_1,\dots q_{p-2})$ and integers $(p,m)$.
\end{enumerate}
In previous sections we have found explicit expressions for ${\GG}_{p}$ in all dimensions for 
$p=(1,2,3)$ (see equations  (\ref{MaxwEvenDim}), (\ref{EvenBX}), (\ref{ODE3form})). For the four--form, we looked only at the ten--dimensional case (\ref{ODE4form}) that gave limited information about 
$\GG_4$:
\bea
\GG_4[x^8]=-\frac{1}{e_2q_1q_2},\quad \GG_4[x^k]=0\ \mbox{for}\ k<8.
\eea
Evaluation of $\GG_p[x^m]$ for all values of parameters $(p,m)$, which would amount to finding the ordinary differential equations (\ref{ODEallFormConj}) for all forms in all even dimensions, is beyond the scope of this paper.

\bigskip

We conclude this section with a brief discussion of the odd--dimensional case. Our results for $p=(1,2,3)$ forms suggest that the ansatz (\ref{pformAnstz}) should modified as
\bea\label{pformAnstzOdd}
&&\hskip -0.5cm 
m^{I_1\mu_1}_{\alpha_1}\dots m^{I_p\mu_p}_{\alpha_p} A_{\mu_1\dots \mu_p}=F_{I_1\dots I_p} 
\left[\prod_{k=1}^p h^{I_k}_{\alpha_k}\right]m^{I_1\mu_1}_{\alpha_1}\dots m^{I_p\mu_p}_{\alpha_p}
\d_{\mu_1}\dots \d_{\mu_p}\Psi\,,\\
&&\hskip -0.5cm m^{I_1\mu_1}_{\alpha_1}\dots m^{I_{p-1}\mu_{p-1}}_{\alpha_{p-1}}n^{\mu_p} A_{\mu_1\dots \mu_p}={\hat F}_{I_1\dots I_{p-1}} 
\left[\prod_{k=1}^{p-1} h^{I_k}_{\alpha_k}\right]m^{I_1\mu_1}_{\alpha_1}\dots 
m^{I_{p-1}\mu_{p-1}}_{\alpha_{p-1}}n^{\mu_p}
\d_{\mu_1}\dots \d_{\mu_p}\Psi\,.\nonumber
\eea
Functions $F_{I_1\dots I_p}$, $h^I_\pm$, and $\Psi$ are still given by (\ref{TotalPformGen}), (\ref{PsiPform}), and ${\hat F}_{I_1\dots I_{p-1}}$ is defined as
\bea\label{3formFFOdd}
{\hat F}_{I_1\dots I_{p-1}}=\Big[\prod^{p-1}_{k<l}[x_{I_k}^2-x_{I_l}^2]\Big]\Big[\prod^{p-1}_{k}x_{I_k}^2\Big]\,.
\eea
Substitution of the ansatz (\ref{pformAnstzOdd}) into the dynamical equation (\ref{EqnCp}) leads to a system of ODEs that generalizes (\ref{OddMPmaster}), (\ref{OddBX}), (\ref{ODE3formOdd}),  (\ref{AptP3odd}):
\bea\label{ODEallOdd}
&&\hskip -1cm
\frac{D_x}{x}\frac{d}{dx}\left[\frac{H_x}{xD_x}X'\right]-\frac{W_x^2H_x X}{x^2}+
F^{(p)}_x\left\{e_2 G_\star+\frac{1-e_2 x^2}{x^2}G_x+i e_3{\GG}_{p}[W_xH_x]\right\}X\nn
&&+\left\{\frac{{\AA}\Omega^2D_x}{x^2}-
\frac{2ie_3H_xW_x(1-e_2 x^2)}{x^2}+D_xP_{n-p-2}[x^2]\right\}X=0\,,\nn
&&\phantom{empty}\\
&&\hskip -1cm
\frac{D_r}{r}\frac{d}{dr}\left[\frac{\Delta}{rD_r}\Phi'\right]+\frac{W_r^2H_r^2}{r^2\Delta}\Phi+
F^{(p)}_r\left\{e_2 G_\star-\frac{1+e_2 r^2}{r^2}G_r+i e_3{\GG}_{p}[W_xH_x]_{x=ir}\right\}\Phi\nn
&&-\left\{\frac{{\AA}\Omega^2D_r}{r^2}-
\frac{2ie_3H_rW_r(1+e_2 r^2)}{r^2}-D_rP_{n-p-2}[-r^2]\right\}\Phi=0\,.\nonumber
\eea 
Functions $\GG$ have the properties (a)--(c) listed on the previous page, and the explicit expressions for  
$\GG_1$, $\GG_2$, $\GG_3$ can be inferred from sections \ref{SecMPMaxw}, \ref{Sec2formOdd}, \ref{Sec3formOdd}.

\bigskip
\noindent
To summarize, in this section we proposed the separable ansatze (\ref{pformAnstz}), (\ref{pformAnstzOdd}) for arbitrary $p$--forms and conjectured the structures (\ref{ODEallFormConj}), (\ref{ODEallOdd}) of the resulting ordinary differential equations. For the special configurations (\ref{OmZeroMain3form}), the derivation of these results is presented in the Appendix \ref{AppA3}, but full justification of equations  (\ref{ODEallFormConj}), (\ref{ODEallOdd}) in the general case for $p>4$ is an interesting open problem.

\section{Extension to the Kerr-(A)dS geometry}
\label{SecAdS}
\renewcommand{\theequation}{7.\arabic{equation}}
\setcounter{equation}{0}

The main motivation for this article comes from the desire to understand the dynamics of $p$--form fluxes on nontrivial stringy backgrounds. The Myers--Perry geometry, which solves equations of motion of supergravity, is a natural example of such a background, so until now we have focused on excitations of this space. 
The equations for scalar and vector fields on this geometry have been solved in the past, and interestingly, success in separation of variables for such excitations extends to rotating black holes with cosmological constant \cite{Kub1,LMaxw}. This turns out to be true for the higher forms as well, but so far we have focused only on vanishing cosmological constant since, as we will explain below, this seems to be the only physically interesting case. 
For completeness, in this section all results are extended to black holes with cosmological constant, even though physical applications of equation (\ref{EqnCp}) on such spaces are not clear.

\bigskip

In certain limits, string theory reduces to supergravities in ten and eleven dimensions, and light excitations of such theories contain various $p$--forms. In particular, the $D=10,11$ supergravities admit solutions of the form 
\bea\label{MPtimesT}
\mbox{MP}_d\times T^{D-d}\,,
\eea  
and in this article we have studied the $p$--form excitations of such spaces. Since the geometry (\ref{MPtimesT}) is a pure metric, linearized equations for all fluxes have the form (\ref{EqnCp}), so all results obtained in previous sections are directly applicable to string theoretic excitations of the geometry  (\ref{MPtimesT}). 

In addition to backgrounds (\ref{MPtimesT}), supergravities also admit solutions of the form\footnote{The best--known examples are $(d,q)=(5,5),(3,3),(2,2)$ for $D=10$ and $(d,q)=(4,7),(7,4)$ for $D=11$.}
\bea\label{MPtimesAdS}
\mbox{AdS}_d\times S^{q}\times T^{D-d-q}\quad \mbox{supported by fluxes}\,.
\eea  
One can try to combine the structures appearing in (\ref{MPtimesT}) and (\ref{MPtimesAdS}) by replacing the AdS factor in the last equation by a black hole with a negative cosmological constant. Such ``Myers--Perry--AdS'' solutions were constructed in the article \cite{GLPP}, and we will refer to them as GLPP geometries,
whose explicit metrics will be written below. Therefore in string theory it is very natural to study excitations of spacetimes 
\bea\label{GLPPtimesAdS}
\mbox{GLPP}_d\times S^{q}\times T^{D-d-q}\quad \mbox{supported by fluxes}\,.
\eea  
Note that, unlike the pure metric (\ref{MPtimesT}), the geometry (\ref{GLPPtimesAdS}) necessarily contains fluxes, which modify the equation (\ref{EqnCp}) for the $p$--forms, making separation of (\ref{EqnCp}) with $p>1$ irrelevant for studying supergravity excitations of (\ref{EqnCp}). Let us explain this in more detail. 

The dynamics of scalars and vectors on the GLPP geometry was analyzed in \cite{Kub1,LMaxw}, where it was shown that the relevant equations separate, just as in the case of the Myers--Perry black hole. 
Furthermore, a 
wave equation for a scalar on the full space (\ref{GLPPtimesAdS}) reduces to a Klein--Gordon equation on GLPP, and similar reduction occurs for Maxwell's equations, so the results of \cite{Kub1,LMaxw} lead to full integrability of dynamical equations for scalar and vector fields on (\ref{GLPPtimesAdS}). Unfortunately, equations of motion for higher forms depend on the structure of background fluxes, and they do not have the universal form (\ref{EqnCp}). For example, there are two two--form excitations $(B^{(2)},C^{(2)})$ of the space
\bea\label{GLPP5}
\mbox{GLPP}_5\times S^{5}\quad \mbox{supported by}\quad 
F_5=L\left[\mbox{vol}_{S^5}+\mbox{vol}_{{GLPP}_5}\right]\,,
\eea
and linearized equations of supergravity mix them \cite{IIBSUGRA,PS}:
\bea\label{PSeqn}
d\star dB^{(2)}=-F_5\wedge dC^{(2)},\quad d\star dC^{(2)}=F_5\wedge dB^{(2)}\,.
\eea
This system is more complicated than a single equation $d\star dA^{(2)}=0$, which appears to be irrelevant for analyzing supergravity excitations of (\ref{GLPP5}). Nevertheless in this section we will study a {\it formal equation} $d\star dA^{(2)}=0$, as well as its generalization (\ref{EqnCp}), in the GLPP geometry and demonstrate full separation of variables, extending the results from the previous section. The analysis of the {\it physical equations}, such as (\ref{PSeqn}) on (\ref{GLPP5}), is an interesting open problem. 

\bigskip

To demonstrate separability of equations (\ref{EqnCp}) in the GLPP geometry, we should begin with giving the explicit form of the metrics. As in the Myers--Perry case, the odd and even dimensions must be treated separately. The geometries were constructed in \cite{GLPP}, and the special frames that generalize (\ref{AllFramesMP}) and (\ref{AllFramesMPOdd}) were found in \cite{Kub1,Kub2,Kub3}. In even dimension one gets
\bea\label{AllFramesAdS}
e_t&=&-\frac{1}{Q_r}\sqrt{\frac{R^2}{FR\Delta}}\left[\d_t-
\sum_k\frac{a_k}{r^2+a_k^2}\d_{\phi_k}\right],\quad 
e_r=Q_r\sqrt{\frac{\Delta}{FR}}\d_r,\nn
e_i&=&-\frac{1}{Q_i}\sqrt{\frac{H_i}{d_i}}\left[\d_t-\sum_k\frac{a_k}{a_k^2-x_i^2}\d_{\phi_k}
\right],\quad e_{x_i}=Q_i\sqrt{\frac{H_i}{d_i}}\d_{x_i}\,.
\eea
The notation is the same as in (\ref{AllFramesMP}), with two new ingredients:
\bea\label{QfactGLPP}
Q_r=\sqrt{1-Lr^2\frac{R}{\Delta}},\quad Q_i=\sqrt{1+Lx_i^2}\,.
\eea
Parameter $L$ introduced in \cite{GLPP} is related to the value of the cosmological constant, and expressions (\ref{AllFramesMP}) are recovered for $L=0$. The counterparts of the null vectors (\ref{Mframes}) are defined by
\bea\label{MframesAdS}
&&m^{(0)}_\pm\equiv \sqrt{FR}(e_r\mp e_t)
=\frac{R}{\sqrt{\Delta}}\left\{\frac{Q_r\Delta}{R}\d_r\pm 
\frac{1}{Q_r}\left[\d_t-
\sum_k\frac{a_k}{r^2+a_k^2}\d_{\phi_k}\right]\right\},\nn
&&m_\pm^{(j)}\equiv\sqrt{d_i}(e_{x_i}\mp e_i)
=\sqrt{{H_j}}\left\{Q_j\d_{x_j}\pm \frac{i}{Q_j}\left[\d_t-\sum_k\frac{a_k}{a_k^2-x_j^2}\d_{\phi_k}
\right]\right\}\,.
\eea
To solve the equation (\ref{EqnCp}) in the even--dimensional GLPP geometry, we introduce the ansatz 
(\ref{pformAnstz}), (\ref{PsiPform}), (\ref{TotalPformGen})--(\ref{EqnForQgen}). This results in the generalization of the ODEs (\ref{ODEallFormConj}),
\bea\label{ODEallAdS}
&&\hskip -1cm
D_x\frac{d}{dx}\left[\frac{Q_x^2H_x}{D_x}X'\right]-\frac{W_x^2H_x}{Q_x^2}X-
F^{(p)}_x\left\{(1-e_2 x^2)\left[G_x-\frac{G_0}{2}\right]+{ie_3 x^2}{\GG}_{p}[W_xH_x]\right\}X\nn
&&+D_xP_{n-p-1}[x^2] X=0,\nn
&&\phantom{42}\\
&&\hskip -1cm
-D_r\frac{d}{dr}\left[\frac{Q_r^2\Delta}{D_r}\Phi'\right]-\frac{W_r^2H_r^2}{Q_r^2\Delta}\Phi-
F^{(p)}_r\left\{(1+e_2 r^2)\left[G_r-\frac{G_0}{2}\right]-{ie_3 r^2}{\GG}_{p}[W_xH_x]_{x=ir}\right\}\Phi\nn
&&+D_xP_{n-p-1}[-r^2] \Phi=0.\nonumber
\eea
Note that the cosmological parameter $L$ appears only in the terms
\bea\label{ModifAdS}
D_x\frac{d}{dx}\left[\frac{Q_x^2H_x}{D_x}X'\right],\quad\frac{W_x^2H_x}{Q_x^2}X\,\quad
\mbox{and}\quad D_r\frac{d}{dr}\left[\frac{Q_r^2\Delta}{D_r}\Phi'\right],\quad
\frac{W_r^2H_r^2}{Q_r^2\Delta}\Phi\,,
\eea
while all other ingredients of equations (\ref{ODEallFormConj}) remain the same as in the Myers--Perry case. 
Modifications (\ref{ModifAdS}) can also be easily implemented in the more explicit equations (\ref{MaxwEvenDim}), (\ref{EvenBX}), (\ref{ODE3form}), (\ref{ODE4form}) for $p=(1,2,3,4)$. 
\bigskip

In odd dimensions, the GLPP metric \cite{GLPP} can be written in terms of separable frames \cite{Kub1,Kub2,Kub3,LMaxw}
\bea\label{FramesAdSOdd}
&&\hskip -1cm e_t=-\frac{1}{Q_r}\sqrt{\frac{R^2}{FR\Delta}}\left[ \d_t
-\sum_k\frac{a_k}{r^2+a_k^2}\d_{\phi_k}\right],\quad e_r=Q_r\sqrt{\frac{\Delta}{FR}}\d_r,\quad
e_{x_i}=Q_i\sqrt{\frac{H_i}{x_i^2 d_i}}\d_{x_i}\nn
&&\hskip -1cm e_i=-\frac{1}{Q_i}\sqrt{\frac{H_i}{x^2_id_i}}\left[\d_t-\sum_k\frac{a_k}{a_k^2-x_i^2}\d_{\phi_k}
\right],\quad 
e_\psi=-\frac{\prod a_i}{r\prod x_k}\left[\d_t-\sum_k\frac{1}{a_k}\d_{\phi_k}
\right],
\eea
which are obtained by inserting the factors $(Q_r,Q_i)$ given by (\ref{QfactGLPP}) into (\ref{AllFramesMPOdd}). The separable ansatz for the $p$--forms ((\ref{pformAnstzOdd}), (\ref{3formFFOdd}), (\ref{TotalPformGen}), (\ref{PsiPform})) is built using the null vectors (\ref{MframesAdS})\footnote{Recall that the expressions for $\Delta$ are different in even and odd dimensions: see equations (\ref{MiscElliptic}) and (\ref{DelktaOdd}).}, as well as $n=e_\psi$. Substitution of the resulting form into the dynamical equation (\ref{EqnCp}) leads to a system of ODEs 
\bea\label{ODEallOddAds}
&&\hskip -1cm
\frac{D_x}{x}\frac{d}{dx}\left[\frac{Q_x^2H_x}{xD_x}X'\right]-\frac{W_x^2H_x X}{x^2Q_x^2}+
F^{(p)}_x\left\{e_2 G_\star+\frac{1-e_2 x^2}{x^2}G_x+i e_3{\GG}_{n,p}[W_xH_x]\right\}X\nn
&&+\left\{\frac{{\AA}\Omega^2D_x}{x^2}-
\frac{2ie_3H_xW_x(1-e_2 x^2)}{x^2}+D_xP_{n-p-2}[x^2]\right\}X=0\,,\nn
&&\phantom{empty}\\
&&\hskip -1cm
\frac{D_r}{r}\frac{d}{dr}\left[\frac{Q_r^2\Delta}{rD_r}\Phi'\right]+\frac{W_r^2H_r^2}{r^2Q_r^2\Delta}\Phi+
F^{(p)}_r\left\{e_2 G_\star-\frac{1+e_2 r^2}{r^2}G_r+i e_3{\GG}_{p}[W_xH_x]_{x=ir}\right\}\Phi\nn
&&-\left\{\frac{{\AA}\Omega^2D_r}{r^2}-
\frac{2ie_3H_rW_r(1+e_2 r^2)}{r^2}-D_rP_{n-p-2}[-r^2]\right\}\Phi=0\,.\nonumber
\eea 
This is a very simple modification of equations (\ref{ODEallOdd}), and similar insertions of the $Q$--factors should be made in the special cases (\ref{OddMPmaster}), (\ref{OddBX}), and (\ref{ODE3formOdd}). 

\bigskip
\noindent
To summarize, in this section all results obtained in our article have been extended to the GLPP geometries, which generalize the Myers--Perry solutions to black holes in the presence of cosmological constant. If would be very interesting to find further extensions to equations like (\ref{PSeqn}), which describe the dynamics of $p$--forms in the spaces with {\it effective} cosmological constants generated by fluxes. Such extensions will provide valuable tools for studying excitations of black holes in string theory.

\section{Discussion}

In this article we have demonstrated separability of the dynamical equations describing all $p$--forms in the backgrounds of the Myers--Perry and GLPP black holes, generalizing the results known for the scalar and vector fields. The most general separable ansatze are given by ((\ref{pformAnstz}), (\ref{TotalPformGen})) in even and by (\ref{pformAnstzOdd})--(\ref{3formFFOdd}) in odd dimensions. The structures of the resulting ordinary differential equations are summarized in (\ref{ODEallFormConj}) and (\ref{ODEallOdd}), and in the special cases of $p=(1,2,3,4)$, the explicit ODEs are given by (\ref{MaxwEvenDim}), (\ref{OddMPmaster}), (\ref{EvenBX}), (\ref{OddBX}), (\ref{ODE3form}), (\ref{ODE3formOdd}), (\ref{ODE4form}).

Our procedure relied on existence of the Killing--Yano tensors, and very restrictive uniqueness theorems for such objects \cite{UniqueKY} imply that it would be hard to extend the methods presented here beyond 
the Myers--Perry and the GLPP geometries, if one focuses only on vacuum solutions of Einstein's equations. 
However, one should be able to incorporate charged black holes since they also admit Killing--Yano tensors \cite{ChLKill}. Unfortunately, equations for interesting excitations of such spacetimes mix various fluxes (see the discussion around formula (\ref{PSeqn})), so we leave exploration of these modes to future work. Another interesting direction involves the study of integrable excitations which are not covered by separable ansatze. 
Although separability of dynamical equations guarantees integrability, the inverse relationship is not true. In particular, there are several examples of gravitational backgrounds where the equation for the scalar field is fully integrable, while there is no separation of variables \cite{LTian}. It would be interesting to see whether a similar phenomenon occurs for the higher forms. Finally, it would also be very interesting to extend the framework introduced this article to gravitational waves.

\section*{Acknowledgments}

This work was supported in part by the DOE grant DE\,-\,SC0017962.

\appendix

\section{Derivation of the separable ansatz}
\renewcommand{\theequation}{A.\arabic{equation}}
\setcounter{equation}{0}
\label{AppA}

This appendix is dedicated to justifying the ansatze (\ref{6DanswCrct}), (\ref{3FrmAnstz}), (\ref{pformAnstz}), (\ref{TotalPformGen}). Specifically, we focus on configurations which do not depend on the cyclic coordinates and determine the relations between various components of the gauge potential and the ``master'' scalar function. The extensions of these relations to configurations with general angular dependence are straightforward, but the intermediate formulas in the derivations are rather cumbersome, 
so  we just write the final answers in the appropriate sections\footnote{For the Maxwell's equations, the detailed derivations were presented in \cite{LMaxw}, and similar analysis is applicable to all forms.}. 

We begin with reviewing the results for electromagnetism by presenting a modification of the analysis performed in \cite{LMaxw} that can be easily extended to higher forms. Then in section \ref{AppA2} we derive the ansatz (\ref{6DanswCrct}) for the two--form, and in section \ref{AppA3} the results are extended to all $p$--form potentials. 

\subsection{Maxwell's equations}
\label{AppA1}

In this subsection we will justify the ansatz (\ref{MaxwAnstz}) for special configurations with
\bea\label{OmMeq0}
\omega=0,\quad n_i=0.
\eea
Although the result  (\ref{MaxwAnstz}) has been already derived in \cite{LMaxw} using a slightly different method, it is the logic presented here that can be easily extended to higher forms, so it is instructive to begin with applying the new method to the Maxwell field before extending it to higher forms in sections \ref{AppA2} and \ref{AppA3}. 

\bigskip

We begin with studying Maxwell's equations in even dimensions, and some peculiarities of the odd--dimensional case will be discussed after equation (\ref{SMaxwAngPolar}). Since the frame $[m^{(I)}_\pm]^\mu$ defined by (\ref{Mframes}) is associated with coordinate $x_I$, a reasonable requirement for separation of variables is\footnote{In (\ref{RadPol1}) and  (\ref{RadPol2}) summation is performed only over index $\mu$.}
\bea\label{RadPol1}
[m^{(I)}_\alpha]^\mu A_\mu=S_{I,\alpha}(x_I)[m^{(I)}_\alpha]^\mu \d_\mu\Psi,\quad \alpha=\pm
\eea
with some unknown functions $(S_{r,\pm},S_{i,\pm})$. For $\Psi$ that does not depend on $(t,\phi_k)$, the relation above simplifies
\bea\label{RadPol2}
[m^{(I)}_\alpha]^\mu A_\mu=S_{I,\alpha}(x_I)[m^{(I)}_\alpha]^I \d_I\Psi\,.
\eea
In the absence of $(t,\phi_k)$--dependence the Maxwell's equations for $(A_r,A_{x_i})$ decouple from the ones for $(A_t,A_{\phi_i})$, so it is convenient to analyze these sectors separately. Similar decoupling will take place for the higher forms discusses in sections \ref{AppA2}, \ref{AppA3}. For the Maxwell field, we will analyze the $(A_r,A_{x_i})$ components in subsection \ref{AppA11} and focus on $(A_t,A_{\phi_i})$ components in section \ref{AppA12}. 

\subsubsection{Non--cyclic components of the gauge field}
\label{AppA11}

Let us consider the Maxwell's equations involving
\bea
A_r\quad\mbox{and}\quad A_a\equiv A_{x_a},\quad a=1\dots n.\nonumber
\eea
Recalling the frames (\ref{Mframes}), we find
\bea\label{RadPol2NonCyc}
A_r=[S_{r,+}+S_{r,-}]\d_r\Psi\equiv S_r\d_r\Psi,\quad 
A_a=[S_{a,+}+S_{a,-}] \d_a\Psi\equiv S_a\d_a\Psi\,.
\eea
The cyclic components of the gauge potential, $(A_t,A_{\phi_k})$, are more complicated, but they do not mix with (\ref{RadPol2NonCyc}) in the Maxwell's equations. In four dimensions, $n=1$ and configurations (\ref{RadPol2NonCyc}) describe a pure gauge (see \cite{LMaxw} for details), therefore here we focus on $n\ge 2$.

Using indices $(i,j,\dots)$ to cover $r$ and $x_a$ and defining a function
\bea
H_r\equiv\Delta=R-Mr,
\eea 
we find the expressions for the gauge potential and the field strength in the sector (\ref{RadPol2NonCyc})
\bea
A_i=S_i\d_i\Psi,\quad F_{ij}=(S_j-S_i)\d_i\d_j \Psi,\quad F^{ij}=\frac{{H_iH_j}}{{d_id_j}}(S_j-S_i)\d_i\d_j \Psi\,.
\eea
Furthermore, evaluating the determinant of the Myers--Perry metric (\ref{MPeven}) in coordinates 
$(r,x_a,t,\phi_a)$,
\bea\label{SqrtGeven}
\sqrt{-g}=\frac{1}{\prod a_k}\left[FR\prod\frac{d_k}{c_k^2}\right]^{1/2}\,,
\eea
we write the $i$--th component of Maxwell's equations as
\bea\label{Sept25}
0=\frac{1}{\sqrt{\prod d_k}}\sum_{j\ne i}\d_j\left[\sqrt{\prod d_k}\frac{{H_iH_j}}{{d_id_j}}(S_j-S_i)\d_i\d_j \Psi\right]
=
\sum_{j\ne i}\frac{H_i}{d_j}\d_j\left[\frac{{H_j}}{{d_i}}(S_j-S_i)\d_j\,\d_i\Psi\right]\,.\nonumber
\eea
Simplifications lead to the final system of equations
\bea\label{RadPolarMaxw}
\sum_{j\ne i} \frac{x_i^2-x_j^2}{d_j}\d_j\left[\frac{{H_j}(S_j-S_i)}{x_i^2-x_j^2}\d_j\,\d_i\Psi\right]=0,\quad 
i=(r,x_a).
\eea
Separable ansatz in $d=2(n+1)$ dimensions implies that function $\Psi$ can be written as
\bea\label{PsiSepApp}
\Psi=X_r(r)\prod X_a(x_a),
\eea
then the set of equations (\ref{RadPolarMaxw}) reduces to a system\footnote{In four 
dimensions, $n=1$, and there is no 
polynomial $P^{(i)}_{n-2}$ in equation (\ref{SemiFinalODEMaxw}). However, as we already mentioned, in this case configuration (\ref{RadPol2NonCyc}) describes a pure gauge, so we are focusing on $n\ge 2$.}
\bea\label{SemiFinalODEMaxw}
\d_j\left[\frac{{H_j}(S_j-S_i)}{x_i^2-x_j^2}\d_jX_j\right]+P^{(j,i)}_{n-2}(x^2_j;x_i)X_j=0,\quad j\ne i.
\eea
Here $P^{(j,i)}_{n-2}(x^2_j;x_i)$ is a polynomial of degree $(n-2)$ in $x^2_j$, and $x_i$--dependence of this function is still undetermined. Substitution of (\ref{SemiFinalODEMaxw}) back into (\ref{RadPolarMaxw}), leads 
to a system of algebraic relations for $P^{(i,j)}_{n-2}$:
\bea\label{MaxwPlnFix1}
\sum_{j\ne i} \frac{x_i^2-x_j^2}{d_j}P^{(j,i)}_{n-2}(x^2_j;x_i)=0.
\eea
The left--hand side of the last relation is a meromorphic function of all $x_j$  for $j\ne i$, then 
analysis of the asymptotic behavior and residues at various poles demonstrates that polynomials $P^{(j,i)}_{n-2}$ must be the same for all values of $j$:
\bea\label{MaxwPlnFix}
P^{(j,i)}_{n-2}(x^2_j;x_i)=P^{(i)}_{n-2}(x^2_j;x_i).
\eea
Once the relation (\ref{MaxwPlnFix}) is satisfied, equation (\ref{MaxwPlnFix1}) becomes an identity.

To summarize, for the separable ansatz (\ref{PsiSepApp}), the Maxwell's equations (\ref{RadPolarMaxw}) are equivalent to the system of ODEs (\ref{SemiFinalODEMaxw}) with restrictions (\ref{MaxwPlnFix}). Furthermore, as in the scalar case (\ref{SeparWaveEvenKerr}), the polynomials $P^{(i)}_{n-2}$ contain some number of separation constants, in particular, setting these constants to zero, one can get a special case of the system (\ref{SemiFinalODEMaxw}) with 
$P^{(i)}_{n-2}=0$:
\bea\label{SemiFinalODEMaxw0}
\d_j\left[\frac{{H_j}(S_j-S_i)}{x_i^2-x_j^2}\d_jX_j\right]=0,\quad j\ne i.
\eea
The current discussion is applicable to $n>2$, then taking $j=0$ and combining equations (\ref{SemiFinalODEMaxw0}) with $i=1,2$, we find
\bea
\frac{x_1^2-x_0^2}{{H_0}(S_0-S_1)}\d_0\left[\frac{{H_0}(S_0-S_1)}{x_1^2-x_0^2}\right]-
\frac{x_2^2-x_0^2}{{H_0}(S_0-S_2)}\d_0\left[\frac{{H_0}(S_0-S_2)}{x_2^2-x_0^2}\right]=0.\nonumber
\eea
In particular, this implies that
\bea\label{tempFuncEqnMaxw}
\d_2\d_0\ln\left[\frac{{H_0}(S_0-S_2)}{x_2^2-x_0^2}\right]=0.
\eea
This relation and its counterparts with other values of $(i,j)$ in (\ref{SemiFinalODEMaxw0}) can be summarized as a system of equations
\bea\label{FuncEqnMaxw}
\frac{S_i-S_j}{x_i^2-x_j^2}=g(x_i)g(x_j).
\eea
with some function $g(x)$. The analogues of the functional equations (\ref{FuncEqnMaxw}) will play the crucial role in derivations of separable ansatze for higher forms, which will be presented in sections \ref{AppA2}, \ref{AppA3}. 

Explicit evaluation of the derivatives in (\ref{tempFuncEqnMaxw}) gives
\bea
\frac{S_0'S_2'}{(S_0-S_2)^2}=\frac{4x_0x_2}{(x_2^2-x_0^2)^2}\,.
\eea
The most general solution of this differential equation  for function $S$ is
\bea\label{SforMaxwell}
{S(x)=\frac{\alpha x^2+\beta}{\gamma x^2+\delta}}
\eea
with arbitrary constants $(\alpha,\beta,\gamma,\delta)$. Although the result (\ref{SforMaxwell}) was derived for the special choice of separation constants corresponding to reduction of (\ref{SemiFinalODEMaxw}) to (\ref{SemiFinalODEMaxw0}), it is clear that function $S$ cannot depend on the separation constants, so relations (\ref{SforMaxwell}) also hold in the general case (\ref{SemiFinalODEMaxw}). Substitution of (\ref{SforMaxwell}) into (\ref{SemiFinalODEMaxw}) with (\ref{MaxwPlnFix}) gives
\bea
\frac{\alpha\delta-\beta\gamma}{\gamma x_i^2+\delta}\d_j\left[\frac{{H_j}}{\gamma x_j^2+\delta}\d_jX_j\right]+P^{(i)}_{n-2}(x^2_j;x_i)X_j=0,\quad j\ne i.
\eea
This equation implies that 
\bea
P^{(i)}_{n-2}(x^2_j;x_i)=\frac{\alpha\delta-\beta\gamma}{\gamma x_i^2+\delta}P_{n-2}(x^2_j),
\eea
where $P_{n-2}(x^2_j)$ is a polynomial with {\it constant} coefficients. 

To summarize, we have demonstrated that the polarization (\ref{RadPol2NonCyc}) is consistent with the separable ansatz (\ref{PsiSepApp}) if and only if $(S_r,S_a)$ are given by (\ref{SforMaxwell}), and 
functions $X_j$ satisfy ordinary differential equations
\bea\label{FinalODEMaxw}
\d_j\left[\frac{{H_j}}{\gamma x_j^2+\delta}\d_jX_j\right]+P_{n-2}(x^2_j)X_j=0.
\eea
Let us now discuss the equations for $(A_t,A_{\phi_i})$.

\subsubsection{Angular components of the gauge field}
\label{AppA12}

The angular components of the gauge field are determined from combinations of (\ref{RadPol2}), which are complementary to (\ref{RadPol2NonCyc}):
\bea\label{AngPol2NonCyc}
([m^{(I)}_+]^\mu-[m^{(I)}_-]^\mu) A_\mu=[S_{I,+}-S_{I,-}][m^{(I)}_+]^I \d_I{\tilde\Psi}.
\eea
For configuration without angular dependence, (\ref{OmMeq0}), functions $\Psi$ and ${\tilde\Psi}$ are independent, but for nontrivial $m_i$ or $\omega$ we expect to have only one ``master function'' $\Psi$, Thus if one views polarizations (\ref{RadPol2NonCyc}) and (\ref{AngPol2NonCyc}) as {\it limits} (\ref{OmMeq0}) of the general solution, then $\Psi$ and ${\tilde\Psi}$ must be the same. In particular, function ${\tilde\Psi}$ must have the form (\ref{PsiSepApp}) with equations (\ref{FinalODEMaxw}) for various ingredients. Rewriting the angular polarizations (\ref{AngPol2NonCyc}) as
\bea\label{MaxwAngulPol}
[m^{(I)}_\pm]^\mu A_\mu=\pm {\tilde S}_I[m^{(I)}_\pm]^I \d_I\Psi,
\eea
and substituting the ansatz (\ref{AngPol2NonCyc}) into Maxwell's equations, we find $(n+1)$ independent equations:
\bea
\MM^\mu=0,\quad\mbox{where}\quad \MM^\mu\equiv\frac{1}{\sqrt{-g}}\d_i[\sqrt{-g}F^{i\mu}],\quad \mu=(t,\phi_i).
\eea 
It is convenient to look at projections of $\MM^\mu$ on various basis vectors, and straightforward calculations give
\bea\label{MaxwProj}
&&[m^{(C)}_+]_\mu \MM^\mu=-\frac{2}{x_C}\d_C\left[\frac{x_C^2}{d_C}\d_C\left[\frac{H_C}{x_C}{\tilde S}_C\d_C
\Psi\right]\right]+
\sum_i\frac{4x_C H_i {\tilde S}_i}{(x_C^2-x_i^2)d_i}\d_C\d_i\Psi\\
&&\quad +\sum_i\frac{4x_i^2}{(x_C^2-x_i^2)d_i}\d_i\left[\frac{H_i}{x_i}{\tilde S}_i\d_i\Psi\right]
+\sum_i\frac{2{\tilde S}_C(x_C^2-x_i^2)}{d_i}\d_i\left[\frac{H_i}{x^2_i-x_C^2}\d_i\d_C\Psi\right]\,.
\nonumber
\eea
We will focus on $n$ scalar equations, 
\bea
[m^{(C)}_+]_\mu \MM^\mu=0,\quad C=1\dots n,\nonumber
\eea
and the projection $[m^{(0)}_+]_\mu \MM^\mu$ can be analyzed in the same way. To use the differential equation (\ref{FinalODEMaxw}), it is convenient to rewrite the derivatives appearing in the last term of (\ref{MaxwProj}) as 
\bea
\d_i\left[\frac{H_i}{x^2_i-x_C^2}\d_i\d_C\Psi\right]=
\left\{\d_i\frac{x_i}{{\tilde S}_i(x_i^2-x_C^2)}\right\}\frac{{\tilde S}_i H_i}{x_i}\d_C\d_i\Psi+
\frac{x_i}{{\tilde S}_i(x_i^2-x_C^2)}\d_i\left[\frac{{\tilde S}_i H_i}{x_i}\d_i\d_C\Psi\right]\,.\nonumber
\eea
Then equation (\ref{MaxwProj}) becomes
\bea\label{MaxwProjOne}
&&[m^{(C)}_+]_\mu \MM^\mu=-\frac{2}{x_C}\d_C\left[\frac{x_C^2}{d_C}\d_C\left[\frac{H_C}{x_C}{\tilde S}_C\d_C
\Psi\right]\right]+
\sum_i\frac{H_i {\tilde S}_i W_{C,i}}{d_i}\d_C\d_i\Psi\nn
&&\quad +\sum_i\frac{4x_i^2}{(x_C^2-x_i^2)d_i}\d_i\left[\frac{H_i}{x_i}{\tilde S}_i\d_i\Psi\right]
-\sum_i\frac{2{\tilde S}_Cx_i}{d_i {\tilde S}_i}\d_i\left[\frac{{\tilde S}_i H_i}{x_i}\d_i\d_C\Psi\right]\,.
\eea
Here we defined
\bea
W_{C,i}=\frac{4x_C}{x_C^2-x_i^2}+\frac{2{\tilde S}_C(x_C^2-x_i^2)}{x_i}\left\{\d_i\frac{x_i}{{\tilde S}_i(x_i^2-x_C^2)}\right\}\,.
\eea
Substituting the separable ansatz (\ref{PsiSepApp}) into equation (\ref{MaxwProjOne}), one arrives at several consistency conditions, in particular, all $W_{C,i}$ must vanish, and all $\d_i$ derivatives must disappear from the last line of (\ref{MaxwProj}). The latter condition leads to differential equations for functions $X_i$:
\bea\label{ODEangul}
\d_i\left[\frac{H_i}{x_i}{\tilde S}_i\d_i\right]\Psi+P^{(i)}[x_i]\Psi=0,
\eea
while equations $W_{C,i}=0$ allow one to determine the functions ${\tilde S}_i$. Specifically, defining $Q_i=\frac{1}{{\tilde S}_i}$, we find
\bea\label{WeqnTemp}
W_{1,2}=\frac{2}{Q_1x_2(x_2^2-x_1^2)}\left[-2x_1 x_2 Q_1+(x_1^2+x_2^2)Q_2+x_2(x_1^2-x_2^2)Q_2'\right]=0
\eea
Treating this relation as an algebraic equation for $Q_1$, we find that 
\bea\label{WeqnTemp1}
Q_i=c_1 x_i+\frac{c_2}{x_i}
\eea
for $i=1$. Equation $W_{2,1}=0$ leads to a similar relation for $Q_2$, and substituting the result into (\ref{WeqnTemp}), we conclude that coefficients $(c_1,c_2)$ can be arbitrary, but they must be the same for all values of $i$. This leads to the expressions for ${\tilde S}_i$ and to a more explicit form of equation (\ref{ODEangul}):
\bea\label{WeqnTemp2}
{\tilde S}_i=\frac{x_i}{c_1 x_i^2+c_2},\quad \d_i\left[\frac{H_i}{c_1 x_i^2+c_2}\d_i\right]\Psi+P^{(i)}[x_i]\Psi=0.
\eea
As expected, for an appropriate choice of coefficients $(c_1,c_2)$ the last relation reproduces the ODE (\ref{FinalODEMaxw}) describing the ``radial'' polarization. This is a consistency check for our construction since (\ref{WeqnTemp2}) and (\ref{FinalODEMaxw}) emerge in the limit of the same equation (\ref{MaxwEvenDim}) when one takes $\omega=n_k=0$.

Let us now determine the properties of function $P^{(i)}[x_i]$ appearing in (\ref{WeqnTemp2}). 
Substitution of relations (\ref{WeqnTemp2}) into expression (\ref{MaxwProj}) gives
\bea\label{MaxwProj1}
[m^{(C)}_+]_\mu \MM^\mu=\frac{2}{x_C}\d_C\left[\frac{x_C^2}{d_C}P^{(C)}
\Psi\right]
-\sum_i\frac{4x_i^2 P^{(i)}}{(x_C^2-x_i^2)d_i}\Psi
+\sum_i\frac{2{\tilde S}_Cx_i P^{(i)}}{d_i {\tilde S}_i}\d_C\Psi\,.
\eea
The coefficient in front of $\d_C\Psi$ vanishes if and only if
\bea\label{RadPolTemp1}
\frac{2x_C}{d_C}\frac{P^{(C)}}{{\tilde S}_C}+\sum_{i\ne C}\frac{2x_i P^{(i)}}{d_i {\tilde S}_i}=0.
\eea
Let us define a new set of functions ${\tilde P}^{(k)}_{n-1}$ by 
\bea
{\tilde P}^{(k)}_{n-1}[x_k^2]\equiv \frac{x_k P^{(k)}}{{\tilde S}_k}\,.
\eea 
Rewriting equation (\ref{RadPolTemp1}) in terms of these functions and recalling the expression (\ref{MiscElliptic}) for $d_k$, we conclude that ${\tilde P}^{(k)}_{n-1}$ must be a polynomial of degree $(n-1)$ in its argument. Furthermore, taking the limit $x_i\rightarrow x_C$ in (\ref{RadPolTemp1}), we find that ${\tilde P}^{(C)}_{n-1}[x^2]={\tilde P}^{(i)}_{n-1}[x^2]$. Therefore, functions ${\tilde P}^{(k)}_{n-1}$  must be the same for all values of $k$, then
\bea\label{AngPolPk}
P^{(k)}=\frac{{\tilde S}_k}{x_k}{\tilde P}_{n-1}[x_k^2]
\eea
Substitution of the last formula into equation (\ref{RadPolTemp1}) leads to a relation
\bea\label{AnguPolynlIdnty}
\frac{{\tilde P}_{n-1}[x^2_C]}{d_C}+\sum_{i\ne C}\frac{{\tilde P}_{n-1}[x^2_i]}{d_i}=0,
\eea
which is identically satisfied for an arbitrary polynomial\footnote{To prove (\ref{AnguPolynlIdnty}), one observes the left--hand side is a meromorphic function of all $x_k$ and shows that residues at all poles, including infinity, vanish. More details can be found in the Appendix E of \cite{LMaxw}.} ${\tilde P}_{n-1}$. Thus functions 
$P^{(k)}$ must have the form (\ref{AngPolPk}), and relation (\ref{MaxwProj1}) reduces to 
\bea\label{MaxwProj2}
&&[m^{(C)}_+]_\mu \MM^\mu=\frac{2}{x_C}\d_C\left[\frac{x_C{\tilde S}_C}{d_C}{\tilde P}_{n-1}[x^2_C]
\right]\Psi
-\sum_i\frac{4x_i{\tilde S}_i {\tilde P}_{n-1}[x^2_i]}{(x_C^2-x_i^2)d_i}\Psi\,.
\eea
To avoid pole at $x_C^2=-c_2/c_1$, we must require that ${\tilde P}_{n-1}[-c_2/c_1]=0$, then relation (\ref{AngPolPk}) can be rewritten as
\bea\label{AngPolPk2}
P^{(k)}={P}_{n-2}[x_k^2],
\eea
where ${\tilde P}_{n-2}$ is an arbitrary polynomial of degree $(n-2)$. 

To summarize, we have demonstrated that the ansatz (\ref{MaxwAngulPol}), (\ref{PsiSepApp}) solves Maxwell's equations if and only if 
\bea\label{WeqnTemp3}
\label{SMaxwAngPolar}
{\tilde S}_i=\frac{x_i}{c_1 x_i^2+c_2},\quad \d_i\left[\frac{H_i}{c_1 x_i^2+c_2}\d_iX_i\right]+
{P}_{n-2}[x_k^2]X_i=0,
\eea
As expected, this ``angular'' polarization and its ``radial'' counterpart (\ref{FinalODEMaxw}) lead to the same system of ODEs for $X_k=(R,X_a)$. Although the derivation of (\ref{WeqnTemp3}) turned out to be 
somewhat tedious, we note that the main results pertaining to the Maxwell field, equations (\ref{FinalODEMaxw}) and (\ref{SforMaxwell}), were obtained using very transparent calculations, and it is this new derivation that will be easily generalized to higher forms in the remaining part of this appendix. 

\bigskip

We conclude the discussion of the Maxwell field by making comments about odd dimensions. Substituting the ansatz (\ref{RadPol2NonCyc}) into Maxwell's equations, we arrive at a counterpart of (\ref{RadPolarMaxw}):
\bea\label{RadPolarMaxwOdd}
\sum_{j\ne i} \frac{x_j(x_i^2-x_j^2)}{d_j}\d_j\left[\frac{{H_j}(S_j-S_i)}{x_j(x_i^2-x_j^2)}\d_j\,\d_i\Psi\right]=0,\quad 
i=(r,x_a).
\eea
Additional factors of $x_j$ are caused by different expressions for the frames and the determinant of the metric. 
The rest of the derivation proceeds as before, leading to (\ref{SforMaxwell}) and the counterpart of (\ref{FinalODEMaxw}):
\bea\label{FinalODEMaxwOdd}
x_j\d_j\left[\frac{{H_j}}{x_j(\gamma x_j^2+\delta)}\d_jX_j\right]+P_{n-2}(x^2_j)X_j=0.
\eea
For the angular polarization we again find (\ref{AngPol2NonCyc}) and the expression for ${\tilde S}_i$ from (\ref{SMaxwAngPolar}). Furthermore, the static configurations (\ref{OmMeq0}) must have $n^\mu A_\mu=0$. 

\bigskip

Although all results of this subsection have already been obtained in \cite{LMaxw}, the new derivation presented here can be easily extended to higher forms, and such generalizations are carried out in the next two subsections.

\subsection{Two--form potential}
\label{AppA2}

Let us now extend the construction presented in the last subsection to the two--form potential. As in the case of Maxwell's field, we will look for separable solutions of linear equations $d\star dA=0$ in the absence of angular dependence. We will mostly focus on the $x$--components of the gauge field, but as we explicitly saw in the last subsection, the resulting equations would apply to other components as well. The verification of this fact is an straightforward extension of the arguments presented in the Appendix \ref{AppA12}, although the intermediate formulas become rather complicated. 

\bigskip
We begin with looking at $x$--components of a two--form field in even dimensions and imposing a natural generalization of the 
ansatz (\ref{RadPol2NonCyc})\footnote{Recall that in this Appendix indices $(i,j,k)$ cover $n$ coordinates $x_a$, as well as $x_0=-ir$.}
\bea\label{RadPolB2}
A_{ij}=S_{ij}(x_i,x_j)\d_i\d_j\Psi.
\eea
Evaluating the field strength,
\bea\label{DefTildS}
F_{ijk}=(S_{jk}-S_{ik}+S_{ij})\d_i\d_j\d_k \Psi\equiv {\cal S}_{ijk}\d_i\d_j\d_k \Psi,\quad 
F^{ijk}=\frac{{H_iH_jH_k}}{{d_id_jd_k}}{\cal S}_{ijk}\d_i\d_j\d_k \Psi,
\eea
and substituting the result into equation $d\star F=0$, we arrive at a counterpart of equation (\ref{Sept25}): 
\bea
0=\frac{1}{\sqrt{\prod d_p}}\sum_j\d_j
\left[\sqrt{\prod d_p}\frac{{H_iH_jH_k}}{{d_id_jd_k}}{\cal S}_{ijk}\d_i\d_j\d_k \Psi\right]
=\sum_j
\frac{H_i H_k}{d_j}\d_j\left[\frac{{H_j}}{{d_i d_k}}{\cal S}_{ijk}\d_j\,\d_i\d_k\Psi\right]\,.\nonumber
\eea
Simplifications of the last relation leads to a system of equations labeled by indices $i$ and $k$:
\bea\label{MaxwEqnB}
\sum_{j\ne i,k} \frac{(x_i^2-x_j^2)(x_k^2-x_j^2)}{d_j}\d_j\left[\frac{{H_j}{\cal S}_{ijk}}{(x_i^2-x_j^2)(x_k^2-x_j^2)}\d_j\d_i\d_k\Psi\right]=0
\eea
Imposing a separable ansatz for function $\Psi$,
\bea\label{PsiSepAppB}
\Psi=X_r(r)\prod X_a(x_a),
\eea
one arrives at a system of differential equations
\bea
\d_j\left[\frac{{H_j}{\cal S}_{ijk}}{(x_i^2-x_j^2)(x_k^2-x_j^2)}\d_jX_j\right]+f^{(j;ik)}(x_j;x_i,x_k)X_j=0.
\nonumber
\eea
The last relation should become a second--order ODE for $X_j(x_j)$, which means that, upon multiplication by an overall function, the coefficients in front of $(\d_j^2 X_j,\d_j X_j,X_j)$ must become functions of $x_j$ only. In particular, this implies the following separation of variables
\bea
\frac{{\cal S}_{ijk}}{(x_i^2-x_j^2)(x_k^2-x_j^2)}=g(x_j)G(x_i,x_k).\nonumber
\eea
Recalling the definition of ${\cal S}_{ijk}$ given by (\ref{DefTildS}), the last relation can be written in a more symmetric form
\bea\label{FunEqnB}
\frac{S_{jk}-S_{ik}+S_{ij}}{(x_i^2-x_j^2)(x_k^2-x_i^2)(x_k^2-x_j^2)}=g(x_i)g(x_j)g(x_k).
\eea
Although a priori we should have used $g^{(i,j,k)}(x)$ instead of $g(x)$ in the last equation, the final ODE (\ref{FinalODEb}) ensures that all $g^{(i,j,k)}(x)$ must be the same, so we avoided unnecessary complications already in (\ref{FunEqnB}). The functional equation (\ref{FunEqnB}), a counterpart of (\ref{FuncEqnMaxw}), is the main consistency condition of the separable ansatz for the two--form potential. 

Substitution of  (\ref{FunEqnB}) into the field equations (\ref{MaxwEqnB}) gives
\bea\label{MaxwEqnBfin}
\sum_{j\ne i,k} \frac{(x_i^2-x_j^2)(x_k^2-x_j^2)}{d_j}\d_j\left[{{H_j}g(x_j)}\d_j\d_i\d_k\Psi\right]=0.
\eea
Imposing the separable ansatz (\ref{PsiSepAppB}) and recalling that $d_j$ is a polynomial of degree $2n$, we arrive at the system of ordinary differential equations
\bea\label{FinalODEb}
\d_j\left[{{H_j}}{g(x_j)}\d_j\right]X_j+P_{n-3}[x^2_j]X_j=0,
\eea
where $P_{n-3}$ is a polynomial of degree $(n-3)$.\footnote{To see this, we observe that, upon multiplication of (\ref{MaxwEqnBfin}) by $\frac{d_m}{(x_i^2-x_m^2)(x_k^2-x_m^2)}$ the terms that do not contain $\d_m$ become polynomials of $(n-3)$ in $x_m^2$.} In particular, for $n=2$ equation (\ref{FinalODEb}) does not contain the last term. Furthermore, for $n=1$, which corresponds to four dimensions, the logic outlined here breaks down, and this degenerate case is discussed in the end of section \ref{Sec2formEven}. 

Let us now solve the functional equations (\ref{FunEqnB}) to determine functions $g(x)$ and $S_{ij}(x_i,x_j)$. Introducing new variables $y_k=x_k^2$ and defining ${\tilde g}_k={\tilde g}(y_k)\equiv g(\sqrt{y_k})$, we can write some integrability conditions of (\ref{FunEqnB}) as (no summations)
\bea\label{IntgrbltB}
\d_{y_i}\d_{y_j}\d_{y_k}\Big[(y_i-y_j)(y_i-y_k)(y_j-y_k){\tilde g}_i{\tilde g}_j{\tilde g}_k\Big]=0.
\eea
Treating this relation as a first--order differential equation for ${\tilde g}_i$, we find
\bea\label{GOfYb}
{\tilde g}(y)=\frac{1}{a y^2+by+c}\,
\eea
with some constants $(a,b,c)$. Substitution of this relation back into (\ref{IntgrbltB}) yields an identity. Let us now look at another integrability condition of (\ref{FunEqnB}):
\bea\label{eqn111}
\d_{y_i}\d_{y_j} S_{ij}&=&\d_{y_i}\d_{y_j}\Big[(y_i-y_j)(y_i-y_k)(y_j-y_k){\tilde g}_i{\tilde g}_j{\tilde g}_k\Big]
\nonumber\\
&=&
\frac{(y_i-y_j)[2a y_iy_j+b(y_i+y_j)+2c]}{[a y_i^2+by_i+c]^2[a y_j^2+by_j+c]^2}
\eea
Integration of this relation gives
\bea
S_{ij}=\frac{y_i-y_j}{a[a y_i^2+by_i+c][a y_j^2+by_j+c]}+A_{ij}(y_i)+B_{ij}(y_j)+C_{ij}.\nonumber
\eea
Substitution of this formula into equation (\ref{FunEqnB}) produces restrictions on $(A_{ij},B_{ij})$:
\bea
B_{ij}(y)=-A_{ij}(y),\quad \d_{y_i}[A_{ij}(y_i)-A_{ik}(y_i)]=0\quad\Rightarrow\quad
A_{ij}(y)=A(y)+{\tilde C}_{ij}.
\eea
This leads to the final formula for the most general $S_{ij}$:
\bea\label{eqn113}
S_{ij}=\frac{y_i-y_j}{a[a y_i^2+by_i+c][a y_j^2+by_j+c]}+A(y_i)-A(y_j)+C_{ij},
\eea
where constants $C_{ij}$ satisfy a system of homogeneous equations
\bea\label{Cidentity}
C_{jk}-C_{ik}+C_{ij}=0.
\eea
With this assignment the functional equation (\ref{FunEqnB}) becomes an identity. 

To summarize, we have found the most general solution of the functional equation (\ref{FunEqnB}):
\bea\label{gSbOne}
g(x_i)=\frac{1}{Q_i},\quad 
S_{ij}=\frac{x^2_i-x^2_j}{aQ_iQ_j}+A(x_i^2)-A(x_j^2)+C_{ij},\quad Q_j=a x_j^4+bx_j^2+c\,.
\eea
Here the set of constants $\{C_{ij}\}$ satisfies the relations (\ref{Cidentity}). Note that these 
constants, as well as $A(y)$, don't contribute to ${\cal S}_{ijk}$ and to the field strength, so the physical degrees of freedom are described by a simpler version of equation (\ref{gSbOne}):
\bea\label{gSbTwo}
S_{ij}=\frac{x^2_i-x^2_j}{aQ_iQ_j}.
\eea
To turn on $(\omega,n_i)$, we would need a slight generalization of the last relation. Using (\ref{gSbTwo}) as an inspiration, we require a multiplicative separation in $S_{ij}$:
\bea\label{eqn117}
S_{ij}=(x^2_i-x^2_j)f(x_i)f(x_j).
\eea
Then consistency with (\ref{gSbOne}),
\bea\label{tempA57}
\d_i\d_j\left[S_{ij}-\frac{x^2_i-x^2_j}{aQ_iQ_j}\right]=0,
\eea
leads to differential equations for function $f$, and the most general solution is\footnote{Strictly speaking, relation (\ref{tempA57}) imposes an additional linear constraint on $(p,q)$. However, this constraint can be eliminated by a rescaling of function $g$ and polynomial $P_{n-3}$ that introduces a multiplicative constant in the right--hand side of equation (\ref{FunEqnB}) without affecting the ODE (\ref{FinalODEb}). Thus we can keep four free parameters $(a,b,c,p,q)$ in (\ref{eqn118}).\label{PageForFoot}}
\bea\label{eqn118}
f(x_j)=\frac{p x_j^2+q}{a x_j^4+bx_j^2+c}\,.
\eea
We conclude that the most general {\it separable} expression for $(g,S_{ij})$ is a special case of (\ref{gSbOne}):
\bea\label{gSbFinal}
\hskip -0.5cm g(x_i)=\frac{1}{Q_i},\ 
S_{ij}=(x^2_i-x^2_j)f(x_i)f(x_j),\ Q_j=a x_j^4+bx_j^2+c,\
f(x_j)=\frac{p x_j^2+q}{Q_j}.
\eea
In section \ref{Sec2form} we demonstrate that this ansatz works for nontrivial $(\omega,n_i)$ as well.

To summarize, we have shown that the most general ``radial'' polarization (\ref{RadPolB2}) with separable function $\Psi$ leads to a system of ODEs (\ref{FinalODEb}). Furthermore, $S_{ij}$ and $g$ must be given by (\ref{gSbFinal}). 

\bigskip

One can also analyze the ``angular'' polarization by modifying the relevant parts of section \ref{AppA12}. This logic of such extension from the Maxwell field to the two--form potential is rather straightforward, but the intermediate formulas are somewhat messy, so here we just quote the result. Starting from an ansatz for the cyclic components\footnote{Recall that for cyclic components $A_{ab}$, indices $(a,b)$ take values $(t,\phi_i)$.} of the gauge field $A_{ab}$,
\bea\label{AngPolB}
\sum_{a,b}[m^{(I)}_\alpha]^a [m^{(J)}_\beta]^b A_{ab}=\alpha\beta{\tilde S}_{IJ}
\sum_{i,j}[m^{(I)}_+]^i[m^{(J)}_+]^j \d_i\d_j\Psi\,,
\eea
with a separable function $\Psi$, (\ref{PsiSepAppB}), and substituting this $A_{ab}$ into the field equations, we find the system of ODEs (\ref{FinalODEb}) with 
\bea\label{AngulPolBans}
g(x_i)=\frac{1}{Q^{(2)}_i},\quad 
{\tilde S}_{IJ}=\frac{(x^2_I-x^2_J)x_Ix_J}{Q^{(2)}_IQ^{(2)}_J},\quad Q^{(2)}_j=a x_j^4+bx_j^2+c.
\eea
Although a priori polynomials $Q_j$ and $Q^{(2)}$ in (\ref{AngulPolBans}) and (\ref{gSbFinal}) have different coefficients, we are interested in obtaining polarization (\ref{RadPolB2}) and (\ref{AngPolB}) as degenerate cases of solutions with nonzero $(\omega,m_i)$. Then these two configurations must be described by the same function $\Psi$, and differential equation (\ref{FinalODEb}) ensures that relations (\ref{AngulPolBans}) and (\ref{gSbFinal}) contain the same parameters $(a,b,c)$.

Finally, for the mixed polarization, we start with the ansatz
\bea
\sum_{\mu}[m^{(I)}_\alpha]^a A_{a j}=
\alpha{\hat S}_{IJ}
\sum_{i}[m^{(I)}_+]^i\d_i\d_j\Psi,
\eea
and determine functions ${\hat S}_{IJ}$ by applying the arguments of section \ref{AppA12}. Once again, the calculations are straightforward but tedious, so we just present the final result:
\bea\label{MixedPolarApp}
\sum_{\mu}[m^{(I)}_\alpha]^a A^{(3)}_{a j}=
\alpha\frac{(x^2_I-x^2_j)x_I({\tilde p}x_i^2+{\tilde q})}{{Q}^{(3)}_I{Q}^{(3)}_j}
\sum_{i}[m^{(I)}_+]^i\d_i\d_j\Psi.
\eea
Function $\Psi$ is given by (\ref{PsiSepAppB}), and its ingredients satisfy a system of differential equations (\ref{FinalODEb}) with
\bea\label{MixedPolBans}
g(x_i)=\frac{1}{Q^{(3)}_i},\quad  Q^{(3)}_j=a x_j^4+bx_j^2+c.
\eea
A priori the coefficients $(a,b,c)$ appearing in the last relation are not the same as parameters in (\ref{AngulPolBans}) and (\ref{gSbFinal}), but if all three polarizations originate from the same ``master function'' $\Psi$, then the differential equation (\ref{FinalODEb}) ensures that all three 
$Q$--factors, (\ref{AngulPolBans}), (\ref{gSbFinal}), (\ref{MixedPolBans}), are built from the same set 
$(a,b,c)$. Combining polarizations (\ref{RadPolB2}), (\ref{AngPolB}), and (\ref{MixedPolarApp}), and requiring the result to be separable, we find
\bea\label{TotalPolB}
&&\sum_{\mu,\nu}[m^{(I)}_\alpha]^\mu [m^{(J)}_\beta]^\nu A_{\mu\nu}=
(x^2_I-x^2_J) h_\alpha^I h_\beta^J\sum_{i,j}[m^{(I)}_+]^i[m^{(J)}_+]^j \d_i\d_j\Psi,\ \\  
&&h_\pm^I=\frac{p x_I^2+q\pm e x_I}{a x_I^4+bx_I^2+c}\,.\nonumber
\eea
This ansatz is the starting point for our discussion of separable two--form potentials in section \ref{Sec2form}. The analysis of odd dimensions follows the same route, and the final result is presented in section \ref{Sec2formOdd}.

\subsection{Higher forms}
\label{AppA3}

Let us now extend the results obtained in the last subsection to $p$--form potentials with $p>2$. We will mostly focus on the ``radial polarization'' (\ref{RadPolB2}) in even dimensions, and comment on the counterparts of (\ref{AngPolB}) and (\ref{MixedPolarApp}) in the end of this subsection. The extension to the odd--dimensional case is straightforward, and it is discussed in sections \ref{Sec3formOdd} and \ref{SecPform}.

The generalization of the ansatz (\ref{RadPolB2}), (\ref{PsiSepAppB}) is
\bea\label{RadPolBp}
A_{i_1\dots i_p}=S_{i_1\dots i_p}(x_{i_1},\dots,x_{i_p})\d_{i_1}\dots \d_{i_p}\Psi,\qquad
\Psi=X_r(r)\prod X_a(x_a).
\eea
Here $S_{i_1\dots i_p}$ is antisymmetric its indices, and we can also introduce an antisymmetric tensor 
${\cal S}_{i_1\dots i_{p+1}}$ by generalizing the definition (\ref{DefTildS}):
\bea\label{DefCalS}
{\cal S}_{i_1\dots i_{p+1}}=S_{i_2\dots i_{p+1}}-S_{i_1i_3\dots i_{p+1}}+
S_{i_1i_2i_4\dots i_{p+1}}-\dots
\eea
Then the field equations, $d\star dA=0$, lead to a counterpart of the system (\ref{MaxwEqnB}):
\bea\label{MaxwEqnBp}
\sum_{j\ne i_1,\dots,i_p} \frac{\prod_k (x_{i_k}^2-x_j^2)}{d_j}\d_j\left[\frac{{H_j}{\cal S}_{ji_1\dots i_p}}{\prod_k (x_{i_k}^2-x_j^2)}\d_j\d_{i_1}\dots \d_{i_p}\Psi\right]=0
\eea
This gives one equation for every set of indices $(i_1,\dots,i_p)$.
Imposing the separable ansatz (\ref{RadPolBp}) for $\Psi$, and repeating manipulations that led to (\ref{FunEqnB}) and (\ref{MaxwEqnBfin}), we can rewrite the system (\ref{MaxwEqnBp}) as 
\bea\label{MaxwEqnBfinP}
\sum_{j\ne i_1,\dots,i_p}  
\frac{\prod_k (x_{i_k}^2-x_j^2)}{d_j}\d_j\left[{{H_j}g(x_j)}\d_j\d_{i_1}\dots \d_{i_p}\Psi\right]=0,
\eea
where tensor ${\cal S}_{i_1\dots i_{p+1}}$ and function $g$ are subject to constraints
\bea\label{FunEqnBp}
{\cal S}_{i_1\dots i_{p+1}}
\prod_{k<l}\frac{1}{(x_{i_k}^2-x_{i_l}^2)}=\prod_{k=1}^{p+1} g(x_{i_k}).
\eea
This functional equation for $S_{i_1\dots i_p}$ and $g$ is a generalization of the constraints 
(\ref {FuncEqnMaxw}) and 
(\ref{FunEqnB}) encountered for the one-- and two--forms. As before, equations (\ref{MaxwEqnBfinP}) with separable function $\Psi$ lead to a system of ODEs
\bea\label{SemiFinalODEbP}
\d_j\left[{{H_j}}{g(x_j)}\d_j\right]X_j+P_{n-p-1}[x^2_j]X_j=0,
\eea
and all these equations contain {\it the same} $P_{n-p-1}$, a polynomial of degree $(n-p-1)$ in its argument. Equations (\ref{SemiFinalODEbP}) apply only for $n\ge (p+1)$, and degenerate cases of small $n$ can be analyzed separately. 

Function $g(x)$ should be determined by solving the functional equation (\ref{FunEqnBp}). The definition (\ref{DefCalS}) implies that 
\bea
\d_{i_1}\dots \d_{i_{p+1}}{\cal S}_{i_1\dots i_{p+1}}=0\quad\mbox{(no summations)},
\eea
leading to an integrability condition for the system (\ref{FunEqnBp}):
\bea\label{FunEqnBpOne}
\d_{i_1}\dots \d_{i_{p+1}}\Big\{
\left[\prod_{k<l}{(x_{i_k}^2-x_{i_l}^2)}\right]\prod_{k=1}^{p+1} g(x_{i_k})\Big\}=0\,
\quad\mbox{(no summations)}.
\eea
Treating the last relation as a first order ODE for the function $g(x_{i_1})$, we find the most general solution of (\ref{FunEqnBpOne}):
\bea\label{gSoln}
g(x_j)=\frac{1}{Q_j},\quad Q_j=\sum_{s=0}^p e_s x_j^{2s}\,.
\eea
Substituting (\ref{gSoln}) into the functional equation (\ref{FunEqnBp}) and solving for $S_{i_1\dots i_p}$, we find\footnote{To derive this, one can follow the steps that led from (\ref{GOfYb}) to (\ref{eqn113}).}
\bea\label{gSbp}
S_{i_1\dots i_p}=\frac{1}{e_p}\Big[\prod_k^p\frac{1}{Q_{i_k}}\Big]\prod^p_{k<l}[x_{i_k}^2-x_{i_l}^2]+
{\hat S}_{i_1\dots i_p}\,,
\eea
where ${\hat S}_{i_1\dots i_p}$ satisfies the homogeneous equation 
\bea\label{HomCHat}
{\hat S}_{i_1\dots i_p}-{\hat S}_{i_1\dots i_{p-1}i_{p+1}}+\dots=0\,.
\eea
Taking various derivatives of this relation and using symmetries of the tensor ${\hat S}$, we find the most general solution of (\ref{HomCHat}): 
\bea\label{eqn134}
{\hat S}_{i_1\dots i_p}(x_{i_1},\dots x_{i_p})=\left[A(x_{i_1},\dots x_{i_{p-1}})+\mbox{permutations}
\right]+
C_{i_1\dots i_p}\,.
\eea
Here $C_{i_1\dots i_p}$ are arbitrary constants satisfying the homogeneous equation
\bea\label{eqn135}
C_{i_1\dots i_p}-C_{i_1\dots i_{p-1}i_{p+1}}+\dots=0\,.
\eea
To summarize, we have shown that the most general solution of the functional equation (\ref{FunEqnBp}) is given by (\ref{gSoln}), (\ref{gSbp}), (\ref{eqn134}), (\ref{eqn135}), and that the system of ODEs 
governing the dynamics of $p$--forms is given by (\ref{SemiFinalODEbP}). 

\bigskip

Using the separable ansatz (\ref{eqn117}) as an inspiration, we focus on a special case of (\ref{gSbp}):
\bea\label{SeparAnstz136}
S_{i_1\dots i_p}=\frac{1}{e_p}\Big[\prod_k^p {f(x_{i_k})}\Big]\prod^p_{k<l}[x_{i_k}^2-x_{i_l}^2]\,.
\eea
We have already found an example that corresponds to ${\hat S}_{i_1\dots i_p}=0$:
\bea
f(x_{i_k})=\frac{1}{Q_{i_k}}\,,
\eea
and now we will determine the most general function $f$ that solves equation (\ref{FunEqnBp}). Since expressions $S_{i_1\dots i_p}$ must have the form (\ref{gSbp}) with (\ref{eqn134}), we conclude that the ansatz (\ref{SeparAnstz136}) solving equation (\ref{FunEqnBp}) must satisfy an integrability condition 
\bea\label{eqn138}
\d_{i_1}\dots\d_{i_p}\left\{\Big[\prod_k^p f(x_{i_k})-\prod_k^p \frac{1}{Q_{i_k}}\Big]\prod^p_{k<l}[x_{i_k}^2-x_{i_l}^2]\right\}=0\,.
\eea
Viewing the left--hand side of this relation as a holomorphic function of $x_{i_1}$, we conclude that the poles of $f(x_{i_1})$ must coincide with zeroes of $Q_{i_1}$. Furthermore, function $f$ must have the 
form
\bea\label{fasTpolyn}
f(x_{i})=\frac{T[x_i^2]}{Q_{i}},
\eea
where $T[x_i^2]$ is a polynomial in $y_{{i}}\equiv(x_{i})^2$. Rewriting equation (\ref{eqn138}) in terms of variables $y_i$, we find
\bea\label{eqn139}
\d_{y_{i_1}}\dots\d_{y_{i_p}}\left\{\Big[\prod_k^p T[y_{i_k}]-1\Big]\Big[\prod_k^p \frac{1}{Q_{i_k}}\Big]\prod^p_{k<l}[y_{i_k}-y_{i_l}]\right\}=0.
\eea
Let us now demonstrate that $T[y_{i_k}]$ must be either a linear polynomial or a constants. 

We begin with factorizing the polynomial $Q$:
\bea\label{eqn141}
Q[y]=e_p\prod_{s=1}^p (y-c_s).
\eea
To simplify the discussion, we assume that all roots of $Q$ are distinct, but the extension to degenerate roots is straightforward. In the vicinity of $y_{i_p}=c_p$, the leading contribution to the left--hand side of (\ref{eqn139}) is proportional to $(y_{i_p}-c_p)^{-2}$, and equation (\ref{eqn139}) can hold only if the residue at this pole vanishes:
\bea\label{eqn142}
\d_{y_{i_1}}\dots\d_{y_{i_{p-1}}}\left\{\Big[T[c_p]\prod_k^{p-1} T[y_{i_k}]-1\Big]\Big[\prod_k^{p-1} \frac{y_{i_k}-c_p}{Q_{i_k}}\Big]\prod^{p-1}_{k<l}[y_{i_k}-y_{i_l}]\right\}=0.
\eea
Rescaling the coefficients of the polynomial $T[y]$ and recalling the factorization (\ref{eqn141}), we can rewrite the relation (\ref{eqn142}) as
\bea\label{eqn143}
\d_{y_{i_1}}\dots\d_{y_{i_{p-1}}}\left\{\Big[\prod_k^{p-1} {\tilde T}[y_{i_k}]-1\Big]\Big[\prod_k^{p-1} \frac{1}{{\tilde Q}_{i_k}}\Big]\prod^{p-1}_{k<l}[y_{i_k}-y_{i_l}]\right\}=0,
\eea
where 
\bea
{\tilde Q}_{i_k}=\prod_{s=1}^{p-1} (y_{i_k}-c_s).
\eea
Relation (\ref{eqn143}) is a {\it necessary condition} for (\ref{eqn139}) to hold, but it has the same form as (\ref{eqn139}) with $p$ replaced by $(p-1)$. Thus polynomial $T$ must have the same degree for $p$-- and $(p-1)$--forms, and by induction, we conclude that the degree of $T$ is the same for all forms. In section \ref{AppA2} we demonstrated that for $p=2$, the polynomial $T$ must be linear (see equation (\ref{eqn118})), so for arbitrary $p$ the degree of $T$ cannot exceed one. It turns out that there are no additional restrictions on $T$: a direct check shows that the functional equation (\ref{FunEqnBp}) is identically satisfied for any  
$S_{i_1\dots i_p}$ given by (\ref{SeparAnstz136}) and (\ref{fasTpolyn}) with an arbitrary linear polynomial\footnote{To keep arbitrary coefficients in $T$, one may have to rescale function $g(x)$, as discussed in the footnote on page \pageref{PageForFoot}.} $T$. To summarize, we have found that the most general separable solution for $S_{i_1\dots i_p}$ is a counterpart of (\ref{gSbFinal})
\bea\label{gSpFormFinal}
g(x_i)=\frac{1}{Q_i},\quad
S_{i_1\dots i_p}=\Big[\prod_j^p \frac{g_0+g_1 x^2_j}{Q_j}\Big]\prod^p_{k<l}[x_{i_k}^2-x_{i_l}^2],
\quad Q_j=\sum_{k=0}^p b_k x_j^{2k},\
\eea
This concludes our discussion of the ``non--cyclic polarization'' (\ref{RadPolBp}) for the $p$--form fields.

\bigskip

In the absence of angular dependence (\ref{OmMeq0}), field equations $d\star dA=0$ split into sectors with a fixed number of cyclic indices. So far we have focused on the non--cyclic components of the gauge field, and we derived the most general separable form of the solution (\ref{gSpFormFinal}). Other sectors can be analyzed in a similar fashion, for example, the result for the sector with one cyclic index reads
\bea\label{eqn146}
\sum_a e^a_{l} A_{i_1\dots i_{p-1}a}=\sum_{i_p}
\Big[\prod^p_{k<l}[x_{i_k}^2-x_{i_l}^2]\Big] \left[\prod_k^{p-1} \frac{T[x_{i_k}^2]}{Q_{i_k}}\right]\frac{x_{i_p}}{Q_{i_p}}
e^{i_p}_{x_l}\,\d_{i_1}\dots\d_{i_p}\Psi,
\eea
The frames $(e^a_{j},e^{i_p}_{x_j})$ appearing here are given by (\ref{AllFramesMP}). Note that for a given value of $j$, the summation in the right--hand side of (\ref{eqn146}) contains only one term. Function $\Psi$ still has the form
\bea
\Psi=X_r(r)\prod X_a(x_a).
\eea
and the dynamical equations $d\star d A=0$ reduce to the system of ODEs (\ref{SemiFinalODEbP}). Analyzing other sectors in the same fashion and combining the results, we arrive at a counterpart of the ansatz (\ref{TotalPolB}):
\bea\label{TotalPolBp}
&&
\sum_{\mu_i}[m^{(I_1)}_{\alpha_1}]^{\mu_1} \dots [m^{(I_p)}_{\alpha_p}]^{\mu_p} A_{\mu_1\dots \mu_p}
\nn
&&\quad=
\Big[\prod^p_{k<l}[x_{i_k}^2-x_{i_l}^2]\Big] \left[\prod_k^p h_{\alpha_k}^{I_k}\right]
\sum_{i_k}[m^{(I_1)}_{+}]^{i_1} \dots [m^{(I_p)}_{+}]^{i_p} \d_{i_1}\dots\d_{i_p}\Psi,\ 
\eea
Coefficients $h_{\alpha_k}^{I_k}$ are rational functions of $x_J$:
\bea\label{TotalPolHHF}  
h_\pm^J=\frac{g_0+g_1 x^2_J\pm g_2x_J}{Q_J}\,,\quad 
Q_j=\sum_{k=0}^p a_k x_j^{2k}\,.
\eea
As demonstrated in sections \ref{Sec3form}, \ref{Sec4form} and \ref{SecPform}, the ansatz (\ref{TotalPolBp}) can be easily extended to configurations with nontrivial $(\omega,n_i)$, although we managed to derive the resulting ODEs only on a case--by--case basis.

\bigskip
\noindent
To summarize, in this Appendix we have derived the most general separable ansatze for arbitrary 
$p$--form fields in the Myers--Perry geometry. Our final result is summarized by 
(\ref{TotalPolBp})--(\ref{TotalPolHHF}), and differential equations produced by this ansatz are discussed in the main body of this article.

\section{Divergence of the two--form potential}
\renewcommand{\theequation}{B.\arabic{equation}}
\setcounter{equation}{0}
\label{AppB}

In the Appendix \ref{AppA} we derived the separable ansatze for the $p$--forms and outlined the derivation of the resulting ordinary differential equations for the ``master field'' $\Psi$. In the presence of angular dependence, the analysis of the equation
\bea\label{EqnPformApB7}
d\star dA^{(p)}=0
\eea
was somewhat messy even in the simplest case of the Maxwell field, i.e., for $p=1$. After such a derivation had been performed in \cite{LMaxw}, the authors of \cite{KubMaxw} {\it observed} that the resulting field 
$A^{(1)}$ satisfied the Lorentz condition
\bea\label{LorMaxwApp}
\nabla_\mu [A^{(1)}]^\mu=0. 
\eea
Then by imposing the ansatz (\ref{MaxwAnstz}) introduced in \cite{LMaxw} along with the Lorentz gauge (\ref{LorMaxwApp}), the authors of \cite{KubMaxw} found an elegant shortcut for arriving at 
(\ref{MaxwEvenDim}), (\ref{OddMPmaster}) which simplified the original derivation of \cite{LMaxw}. In this Appendix we explain the origin of this shortcut and demonstrate that, unfortunately, it cannot be directly extended to higher forms. 

\bigskip

Let us begin with the discussion of the one--form $A\equiv A^{(1)}$ in even dimensions. The ansatz (\ref{MaxwAnstz}),
\bea\label{MaxwAnsApp}
[m^{(I)}_\pm]^\mu A_\mu=\mp\frac{i}{x_I\pm\mu}[m^{(I)}_\pm]^\mu\d_\mu \Psi,\quad x_0=-ir,
\eea
fully fixes the gauge, so one cannot impose the Lorentz condition (\ref{LorMaxwApp}) {\it in addition} to 
(\ref{MaxwAnsApp}). However, as we will now demonstrate, any separable solution (\ref{MaxwAnsApp}) of the Maxwell's equations $d\star dA=0$ {\it automatically} satisfies the condition (\ref{MaxwAnsApp}). We will later see that this property does not extend to higher forms, so the Lorentz gauge 
\bea\label{LorntzHigh}
\nabla_\mu [A^{(p)}]^{\mu\nu_1\dots \nu_{p-1}}=0
\eea
is inconsistent with the dynamical equations (\ref{EqnPformApB7}) for the separable configurations (\ref{pformAnstz})--(\ref{PsiPform}) with $p>1$. To avoid unnecessary complications, we will focus on a  special class of solutions
\bea\label{PsiBfieldAppB}
\Psi=E\Phi(r)\left[\prod X_i(x_i)\right],\qquad E=1\,,
\eea
although the conclusion about satisfaction or failure of the Lorentz gauge holds in the presence of $(t,\phi_i)$ dependence as well.

In the special case (\ref{PsiBfieldAppB}), the non--cyclic components of the Maxwell field (\ref{MaxwAnsApp}) become\footnote{As in the Appendix \ref{AppA}, here we use the notation $A_j=(A_r,A_{x_a})$.}
\bea
A_j=\frac{1}{2}\left[-\frac{i}{x_j+\mu}+\frac{i}{x_j-\mu}\right]\d_j \Psi=\frac{i\mu}{x_j^2-\mu^2}
\d_j \Psi\,.
\eea
Substitution into the left--hand side of (\ref{LorMaxwApp}) gives
\bea\label{tempJun27}
\nabla_\mu A^\mu=\frac{1}{\sqrt{-g}}\d_j\left[\sqrt{-g}g^{jk}A_k\right]=
\frac{1}{FR}\d_r\left[FR\frac{\Delta}{FR}A_r\right]+
\sum_{j=1}^n\frac{1}{d_j}\d_j\left[d_j\frac{H_j}{d_j}A_j\right].
\eea
Here we used the expressions for $g^{ij}$ from (\ref{MPevenInFrames}), as well as equation (\ref{SqrtGeven}) for the determinant of the metric. As we have shown in the Appendix \ref{AppA}, the Maxwell's equations for the separable configurations (\ref{MaxwAnsApp}) are equivalent to the system of ODEs (\ref{MaxwEvenDim}), then in the present case $\omega=n_i=0$ we find (no summation)
\bea
&&\hskip -0.9cm\d_j\left[d_j\frac{H_j}{d_j}A_j\right]=
\frac{\d}{\d x_j}\left[\frac{i\mu H_j}{x_j^2-\mu^2}\frac{\d\Psi}{\d x_j}\right]=
-\frac{i\mu}{\mu^2}\frac{\d}{\d x_j}\left[\frac{H_j}{D_j}\frac{\d\Psi}{\d x_j}\right]=
\frac{i}{\mu}P_{n-2}[x_j^2]\Psi,\quad j=1,\dots,n;\nn
&&\hskip -0.9cm\d_r\left[FR\frac{\Delta}{FR}A_r\right]=
\frac{\d}{\d r}\left[\frac{i\mu \Delta}{r^2+\mu^2}\frac{\d\Psi}{\d r}\right]=
\frac{i\mu}{\mu^2}\frac{\d}{\d r}\left[\frac{\Delta}{D_r}\frac{\d\Psi}{\d r}\right]=
\frac{i}{\mu}P_{n-2}[-r^2]\Psi.\nonumber
\eea
Substitution into (\ref{tempJun27}) gives
\bea
\nabla_\mu A^\mu=\frac{i}{\mu}\left[\frac{P_{n-2}[-r^2]}{FR}+\sum_{j=1}^n\frac{P_{n-2}[x_j^2]}{d_j}\right]\,.
\nonumber
\eea
As we saw in the Appendix \ref{AppA11}, the right-hand side of the last equation vanishes identically, so we have shown that any separable solution (\ref{MaxwAnsApp}), (\ref{PsiBfieldAppB}) of the Maxwell's equations satisfies the Lorentz gauge
\bea
\nabla_\mu A^\mu=0.\nonumber
\eea
The conclusion holds even for a more general function $E$ given in (\ref{WaveEqnTwo}), but the proof requires longer algebraic manipulations, and we will not present them here.

\bigskip

Let us now look at the separable two--form potential (\ref{6DanswCrct}):
\bea
m^{I\mu}_\alpha m^{J\nu}_\beta A_{\mu\nu}=
(x_I^2-x_J^2)\, h^I_\alpha h^J_\beta\, m^{I\mu}_\alpha m^{J\nu}_\beta \d_\mu\d_\nu\Psi,
\quad
h^I_\pm=\frac{1}{e_1+e_2 x_I^2\pm ie_3 x_I}\,.
\eea
Once again we will focus on the special configurations whose ``master function'' has the form (\ref{PsiBfieldAppB}). The non--cyclic components of the gauge field are given by
\bea
A_{jk}=(x_j^2-x_k^2)f_jf_k\d_j\d_k\Psi,\quad f_j=\frac{e_1+e_2 x_j^2}{(e_1+e_2 x_j^2)^2+e^2_3 x_j^2}\,.
\eea
Following the steps used to evaluate $\nabla_\mu A^\mu$, we find the non--cyclic components of the divergence of the two--form:
\bea
&&\hskip -1cm\frac{1}{\sqrt{-g}}\d_j\left[\sqrt{-g}g^{jj'}g^{kk'}A_{j'k'}\right]=
\frac{1}{FR}\d_r\left[FR\frac{\Delta}{FR}\frac{H_k}{d_k}A_{rk}\right]+
\sum_{j=1}^n\frac{1}{d_j}\d_j\left[d_j\frac{H_j}{d_j}\frac{H_k}{d_k}A_{jk}\right]\nonumber\\
&&=\frac{(-r^2-x_k^2)f_kH_k}{d_kFR}\d_r\left[f_r{\Delta}\d_r\right]\d_k\Psi+
\sum_{j=1}^n\frac{(x_j^2-x_k^2)f_k H_k}{d_jd_k}\d_j\left[{H_j f_j}\d_j\right]\d_k\Psi\nn
&&=\frac{2e_2r(r^2+x_k^2)f_kH_k\Delta}{d_kFR D_r}\d_r\d_k\Psi+
\sum_{j=1}^n\frac{2e_2 x_j(x_j^2-x_k^2)f_k H_k H_j}{d_jd_k D_k}\d_j\d_k\Psi\,.\nonumber
\eea
Here we used the ODEs (\ref{EvenBX}) for function $\Psi$, as well as identities (\ref{RadPolarMaxw})--(\ref{SemiFinalODEMaxw}) for the polynomial $P_{n-2}$. The last equation can be summarized as
\bea\label{Lorn2nc}
\nabla_j A^{jk}=\frac{2e_2r(r^2+x_k^2)f_kH_k\Delta}{d_kFR D_r}\d_r\d_k\Psi+
\sum_{j=1}^n\frac{2e_2 x_j(x_j^2-x_k^2)f_k H_k H_j}{d_jd_k D_k}\d_j\d_k\Psi\,.
\eea
Although this expression is very compact, it is clear that the two--form does not satisfy the Lorentz condition (\ref{LorntzHigh}). The cyclic components of the divergence can be computed in a similar fashion, and the result reads
\bea\label{Lorn2cyc}
\nabla_j A^{jb}=\sum_{j<k} 
\frac{2e_2e_3 a_b x_j x_k (x_j^2-x_k^2)^2H_jH_k}{(a_b^2-x_j^2)(a_b^2-x_k^2)D_jD_k d_jd_k}\d_j\d_k\Psi
\eea
Once again, this expression does not vanish.

The expressions (\ref{Lorn2nc}), (\ref{Lorn2cyc}) can be extended to configurations with nontrivial $(t,\phi_i)$ dependence and generalized to higher forms, and although the explicit formulas are more complicated, the conclusion remains the same: the separable configurations (\ref{pformAnstz}) satisfying the equations of motion (\ref{EqnCp}) do not belong to the Lorentz gauge (\ref{LorntzHigh}) for any $p>1$. Thus existence of an elegant shortcut for deriving the ODEs (\ref{MaxwEvenDim}) proposed in \cite{KubMaxw} seems to be a peculiarity associated with Maxwell's equations, and for higher forms one has to follow the longer route discussed in the Appendix \ref{AppA}.

\end{document}